\documentclass[aps,pra,amsmath,amssymb,twocolumn,superscriptaddress]{revtex4-1}
\usepackage{graphicx}
\usepackage{epstopdf}
\usepackage{dcolumn}
\usepackage{bm}
\usepackage{natbib}
\usepackage{braket}

\begin{document}

\preprint{APS/123-QED}

\title{Quantum error correction failure distributions: comparison of coherent and stochastic error models}

\author{Jeff P. Barnes}
\email{jeff.barnes@jhuapl.edu}
\affiliation{Johns Hopkins University Applied Physics Laboratory \\11100 Johns Hopkins Rd., Laurel, MD, 20723, USA}
\author{Colin J. Trout}
\affiliation{Johns Hopkins University Applied Physics Laboratory \\11100 Johns Hopkins Rd., Laurel, MD, 20723, USA}
\affiliation{Schools of Chemistry and Biochemistry, Computational Science and Engineering, and Physics\\ Georgia Institute of Technology, Atlanta, GA 30332, USA}
\author{Dennis G. Lucarelli}
\affiliation{Johns Hopkins University Applied Physics Laboratory \\11100 Johns Hopkins Rd., Laurel, MD, 20723, USA}
\author{B.D. Clader}
\affiliation{Johns Hopkins University Applied Physics Laboratory \\11100 Johns Hopkins Rd., Laurel, MD, 20723, USA}


\begin{abstract}
We compare failure distributions of quantum error correction circuits for stochastic errors and coherent errors. We utilize a fully coherent simulation of a fault tolerant quantum error correcting circuit for a $d=3$ Steane and surface code. We find that the output distributions are markedly different for the two error models, showing that no simple mapping between the two error models exists. Coherent errors create very broad and heavy-tailed failure distributions. This suggests that they are susceptible to outlier events and that mean statistics, such as pseudo-threshold estimates, may not provide the key figure of merit. This provides further statistical insight into why coherent errors can be so harmful for quantum error correction. These output probability distributions may also provide a useful metric that can be utilized when optimizing quantum error correcting codes and decoding procedures for purely coherent errors.
\end{abstract}

\maketitle

\section{
\label{sec:introduction}
Introduction
}
The theory of fault tolerance and the associated threshold theorem show that given an error rate on physical qubits below some threshold, one can perform a quantum computation with arbitrary accuracy with manageable overhead due to error correction (see e.g., \cite{shor1995, steane1996, calderbank1996, aharonov1997, knill1998, aliferis2006, aharonov2008, gottesman2009intro, RevModPhys.87.307}). It is well known that the Pauli matrices form a basis for any arbitrary qubit state. Therefore any arbitrary unitary evolution can be written as a linear combination of Pauli matrices. Thus, it is sufficient to correct only for these types of errors in quantum error correction \cite{nielsen2010}. 

This fact often leads to a logical leap that is not always justified. In particular, in numerical simulations of quantum error correction routines, it is standard practice to only consider stochastic Pauli errors. In some cases such a model is valid. For example, in models of open quantum systems \cite{haake2001,breuer2002,weiss2012}, the system environment coupling causes the environment to constantly ``measure'' the qubits. This can cause qubit depolarization, which can be modeled as the mapping of the qubit density matrix
\begin{equation}
  \rho \rightarrow
  (1-p_1-p_2-p_3)\rho +
  p_1 \mathbf{X}\rho\mathbf{X}
  + p_2 \mathbf{Y}\rho\mathbf{Y}
  + p_3 \mathbf{Z}\rho\mathbf{Z},
  \label{eq:pauli error model}
\end{equation}
where the $p_j$ terms are classical probabilities. Therefore a completely acceptable interpretation of Eq. \eqref{eq:pauli error model} is one where with probability $(1-p_1-p_2-p_3)$ no error happens to the wavefunction, and with probability $p_j$ a Pauli $\mathbf{X}$, $\mathbf{Y}$, or $\mathbf{Z}$ error occurs for $j=\{1,2,3\}$ respectively. This error model is referred to by a variety of different names in the literature. Here we refer to it as the ``stochastic Pauli error model". More generally, an error model can be generated with a larger set of gates than just Pauli gates, but where error gates are randomly inserted in a quantum circuit with classical probabilities. The key characteristic of all these stochastic error models is that the various error configurations all arise from classical probabilities, resulting in no interfering pathways between them.

Coherent errors, which can arise by over- or under-rotations of a quantum gate, were often called ``inaccuracies" in the early literature on quantum error correction \cite{knill1996, shor1997}. They cannot map to any stochastic error model as they result in a fully coherent evolution with multiple interfering pathways. Proving the existence of threshold in the presence of coherent errors thus requires one to consider the fully coherent sum of interfering fault paths (see, e.g., Sec. 8 in Ref. \cite{aharonov2008}). The impact of coherent errors has received renewed attention due to recent results confirming that such errors can be a more severe hurdle for fault-tolerance than stochastic errors \cite{sander2016, ExpFTThresh}.

A more thorough, physics-based justification for when a stochastic error model is an appropriate approximation requires one to consider the entire system-bath interaction. Only when the bath state, entangled with a given error, is orthogonal to all other bath states is the stochastic error model truly appropriate \cite{knill1996, knill1998}. Despite these fairly well-known results from the early literature on quantum error correction, the use of stochastic error models is still widely used in numerical simulations to calculate thresholds, including many results that have examined the accuracy of approximating various error channels by stochastic errors \cite{gutierrez2013, geller2013, puzzuoli2014, witzel2014, darmawan2016tensor, gutierrez2016errors}. Of course, a significant reason for using this error model this is that these types of errors can be efficiently simulated classically via the Gottesman-Knill theorem \cite{gottesman1998heisenberg, gottesman2004} if one restricts the set of gates to the Clifford group or a subset (often just Pauli operators are used). 

Here, we report that a stochastic error model, has a drastically different logical failure distribution than a failure distribution resulting from coherent errors. To show this, we compare the output failure distributions of a quantum error correction (QEC) memory circuit correcting stochastic Pauli errors to coherent errors. We utilize a numerical simulation of the entire encoded wavefunction, and we examine the failure distribution of the logical state for the Steane [[7,1,3]] code and a distance 3 surface code. We devise failure metrics that avoid the need to implement a final round of perfect error correction and decoding, commonly used when studying the failure characteristics of a logical channel \cite{Rahn2002, gutierrez2013, darmawan2016tensor, gutierrez2016errors}. Our results show that the output failure distributions are markedly different for the two error models, with coherent errors lead to very broad failure distributions. This suggests that outlier events are much more likely to occur for coherent errors than for stochastic errors, which may provide insight into why they can be so harmful to QEC \cite{sander2016, ExpFTThresh}.

\section{Numerical Simulation Approach}


We numerically simulate two common QEC $d=3$ codes in this paper: the Steane [[7,1,3]] code \cite{steane1996, nielsen2010}, and the titled-17 surface code \cite{OptimalResourcesTop, SCLattSurgery,FTQCAnyons}. Insertions of errors in the stochastic Pauli (SP) error model are treated as unitary qubit gates, so for both the SP error model and pulse-area error model model the evolution is purely unitary in our simulator. The goal of the report is to determine and compare the output failure probability distributions between these two error models.

Each run starts with a state of Eq.~(\ref{eq:starting state}):
\begin{equation}
  \label{eq:starting state}
  |\Psi_0 \rangle =
  ( \cos\theta\, |0_L\rangle + \sin\theta\,e^{i\phi} |1_L\rangle )\otimes |0000\rangle
\end{equation}
Here,  $\theta = \pi\, \mathrm{rand()}$ and $\phi = 2\pi\, \mathrm{rand()}$,
with rand() referring to a uniform pseudo-random number. This covers the Bloch sphere, though not necessarily uniformly.

\subsection{Quantum Operations}


For a single qubit, the most general evolution is given by an interaction Hamiltonian
\begin{equation}
  \mathbf{H} = \hbar \omega \left(
  u_x \mathbf{X} + u_y \mathbf{Y} + u_z \mathbf{Z}
  \right)
\end{equation}
where $\mathbf{X}$, $\mathbf{Y}$ and $\mathbf{Z}$ are Pauli operators \cite{nielsen2010},
the rotation axis is $(u_x,u_y,u_z)$,
and couplings with field amplitudes are absorbed into $\omega$.
Evolution is given by $|\Psi\rangle \rightarrow \mathbf{U}|\Psi\rangle$,
with the propagator:
\begin{align}\label{eq:single qubit gate}
  \mathbf{U}
 & = \exp( -i \mathbf{H} t / \hbar ) \\ \nonumber
&  = \cos(\omega t)\mathbf{1} -i\sin(\omega t)
  \left( u_x \mathbf{X} + u_y \mathbf{Y} + u_z \mathbf{Z} \right).
\end{align}

In these simulations, the {\it shortest} path
in Hilbert space is used to evolve a gate.
Thus, the Hadamard gate $\mathbf{W} = ( \mathbf{X} - \mathbf{Z} )/\sqrt{2}$
is created with an axis $(1,0,-1)/\sqrt{2}$ and a duration $t = \pi/(2\omega)$.


The CNOT($n\rightarrow m$) gate is implemented
using the Hamiltonian:
\begin{equation}
  \mathbf{H} = \hbar \omega \big( 
  \mathbf{1}-|1_n\rangle \langle 1_n| +
  |1_n\rangle \langle 1_n| \otimes \mathbf{X}_m
  \big).
  \label{eq:CNOT Hamiltonian}
\end{equation}
The projectors ensure that no phase difference accumulates for $|0_n\rangle$.
This same structure extends to more than one controlling qubit, such as the Toffoli gate.
For the controlled-Z gate, replace $\mathbf{X}$ with $\mathbf{Z}$.


A measurement of qubit $n$ s implemented as
\begin{equation}
  |\Psi\rangle \rightarrow \left\{
  \begin{array}{cc}
    |0_n\rangle \langle 0_n |\Psi \rangle / \| \langle 0_n|\Psi\rangle \| & \mathrm{rand()} \le
    \left| \langle 0_n | \Psi \rangle \right|^2 \\
    |1_n\rangle \langle 1_n |\Psi \rangle / \| \langle 1_n|\Psi\rangle \| & \mathrm{otherwise.} \\
  \end{array}
  \right.
  \label{eq:measure definition}
\end{equation}
Measurements are instantaneous, and their records are perfect. These are standard assumptions in QEC.
%
%

%
%
%
\subsection{
\label{sec:control errors}
Error Models}

\subsubsection{
  \label{sec:pauli error model}Pauli Error Model}

A stochastic Pauli error model is implemented by selecting a set of fault locations and an error rate, $p$.
A detailed discussion can be found in \cite{cross2009}.
The choice here is conservative:  errors are restricted to being only
on the codeword qubits, and only before each syndrome measurement. While conservative, this is not unrealistic for this error model as the Pauli errors can be efficiently commuted through a QEC circuit. When the code encounters a fault location, it randomly applies an $\mathbf{X}$ and $\mathbf{Z}$ error each with probability $p$.
Since $\mathbf{X}\mathbf{Z} = -i\mathbf{Y}$, all Pauli errors can occur.

\subsubsection{
  \label{sec:coherent control errors}
  Pulse-Area Error Model}

For coherent errors, we utilize an error model that we call the pulse-area error model. Coherent manipulations of quantum systems almost universally involve the application of an electromagnetic field, and these are imprecise. Static fields are not independent degrees of freedom, and their fluctuations are tied back to their sources \cite{cohentannoudji1989,heitler1954}. Free fields, which include RF pulses in magnetic resonance \cite{ernst1987} and laser pulses in time domain optical spectroscopy \cite{yariv1996}, are usually well represented by coherent states \cite{mandel1995}, and their quantum back-action is quantifiable \cite{barnes1999}. Commonly, though, these terms can be neglected, and the field amplitudes are $c$-functions in the Hamiltonian that describes a qubit gate. A perfect gate requires the integral of the field amplitude over time (in a rotating frame) to be a fixed angle. The pulse-area error model puts a distribution on this angle. The mean is zero, since systematic errors can be removed with calibration \cite{merrill2014}.

The pulse-area error model consists of replacing
$\omega$ in Eq.~(\ref{eq:single qubit gate})
with $\omega(1 + \sigma r)$, with $-1 < r < 1$ uniformly
and independently per gate.
\begin{equation}
  \exp \left(
  -\frac{i}{\hbar} [ \hbar \omega \mathbf{G} ] \frac{\pi}{2\omega}
  \right)
  \rightarrow
  \exp \left(
  -\frac{i}{\hbar}
  [ \hbar ( 1 + \sigma r ) \mathbf{G} ]
  \frac{\pi}{2}
  \right)
  \label{eq:sigma definition}
\end{equation}
All gates in the QEC circuit have $\omega = 1.0$ and
duration $\pi/2$.  Thus, $\sigma$ is related to the fractional jitter
in the field amplitudes.  For example, 
for a commercially available pulsed laser sources,
$\sigma \approx 0.005$ \cite{coherent2016}.

%
%
%

%
%
%
\section{
  \label{sec:metrics}
  Metric for QEC Failure}

This section derives a metric to quantify QEC circuit failure,
but adapted to a wavefunction simulation. The prescription here is for the Steane [[7,1,3]] code with a few special considerations required for the surface code noted in Sec. \ref{sec:space_degeneracy}. The focus here is on developing a metric for failure that does not require any perfect measurements, QEC rounds, or decoding, but rather a metric that can be applied to the wavefunction itself.
 
 \subsection{Steane Failure Metric}
The following definition, from \cite{aliferis2006},
is considered a standard:
a QEC circuit that takes any input state with weight $\le 1$ errors
and outputs a state with weight $\le 1$ errors has succeeded;
otherwise it has failed.
A weight-1 error is defined as having a single Pauli error applied
to the state, a weight-2 error has two errors, and so on.
The goal is to estimate the probability of failure $P_\mathrm{fail}(p)$
over many trial runs at fixed $p$.
The concatenation threshold is based upon the bound
$P_\mathrm{fail}(p) < p$
(but there are subtle corrections, see \cite{aliferis2006}).

This suggests measuring $P_\mathrm{fail}(p)$ as the portion of
$|\Psi(t)\rangle$ that projects into the space of all weight $\ge 2$ errors.
There are several possibilities for this space.
Consider using operators, as in a CHP simulation \cite{gottesman2004}.
For example, the [[3,1,3]] bit flip correcting code
is constructed as in Eq.~(\ref{eq:shor CHP example}),
using $T$ and $C$ to denote Toffoli and controlled-not gates.
The $P$ is a projector to $|0\rangle$ that enforces the requirement of fresh ancilla.
Any bit-flip error between the encoding and repair steps,
by simple matrix multiplication, is seen to never act on qubit 1.
This proves the QEC circuit protects the qubit.
The point is that $|\Psi_0 \rangle$ does not appear
anywhere in this proof.  This suggests the QEC failure space for a $d=3$ code
is built up from the logical codeword space,
as Eq.~(\ref{eq:logical subspaces}).
\begin{widetext}
\begin{eqnarray}
  \overbrace{T(2,3,1)C(1,2)C(1,3)}^{\mathrm{repair}}
  \overbrace{X(1)}^{\mathrm{error}}
  \overbrace{C(1,3)C(1,2)}^{\mathrm{encode}}
  \overbrace{P(2,3)}^{\mathrm{ancilla}}
  &=& X(2)X(3)P(2,3) \nonumber \\
  T(2,3,1)C(1,2)C(1,3) \: X(2) \: C(1,3)C(1,2) \: P(2,3)
  &=& X(2)P(2,3) \nonumber \\
  T(2,3,1)C(1,2)C(1,3) \: X(3) \: C(1,3)C(1,2) \: P(2,3)
  &=& X(3)P(2,3)
  \label{eq:shor CHP example}
\end{eqnarray}
\begin{eqnarray}
  \mathcal{S}_L &=&
  \left\{ |0_L\rangle,\; |1_L\rangle \right\}
  \nonumber \\
  \mathcal{S}_{L+1} &=&
  \mathcal{S}_L \cup
  \left\{
  \mathbf{X}_q|0_L\rangle, \;
  \mathbf{X}_q|1_L\rangle, \;
  \mathbf{Y}_q|0_L\rangle, \;
  \mathbf{Y}_q|1_L\rangle, \;
  \mathbf{Z}_q|0_L\rangle, \;
  \mathbf{Z}_q|1_L\rangle
  \right\}
  \nonumber \\
  \mathcal{S}_{L+2} &=&
  \mathcal{S}_L \cup
  \left\{
  \mathbf{X}_q|0_L\rangle, \;
  \mathbf{X}_q|1_L\rangle, \;
  \mathbf{Z}_q|0_L\rangle, \;
  \mathbf{Z}_q|1_L\rangle, \;
  \mathbf{X}_q\mathbf{Z}_{q^\prime}|0_L\rangle, \;
  \mathbf{X}_q\mathbf{Z}_{q^\prime}|1_L\rangle
  \right\}  \;\;\; \forall q,q^\prime
  \label{eq:logical subspaces}
\end{eqnarray}
\end{widetext}
These kets form orthonormal states,
with $|\mathcal{S}_L| = 2$, $|\mathcal{S}_{L+1}| = 44$,
and $|\mathcal{S}_{L+2}| = 128$ for the particular case of Steane's code.
Further, since $\mathbf{X}_q\mathbf{Z}_q = -i \mathbf{Y}_q$,
one can show $\mathcal{S}_L \subset \mathcal{S}_{L+1} \subset \mathcal{S}_{L+2}$.
But note that the total Hilbert space of Steane's code has $2^7 = 128$ dimensions.
Thus, $\mathcal{S}_{L+2}$ completely spans it.
Since no state could ever leave $\mathcal{S}_{L+2}$,
it cannot represent a QEC failure criteria,
but $\mathcal{S}_{L+1}$ can.

From the vantage point of Monte Carlo wavefunction simulations,
it seems more natural to check if two or more Pauli operators
have acted on the starting state $|\Psi_0 \rangle$ for a $d=3$ code.
These spaces are built up as
\begin{eqnarray}
  \mathcal{S}_\psi &=&
  \left\{ |\Psi_0 \rangle \right\}
  \nonumber \\
  \mathcal{S}_{\psi+1} &=&
  \mathcal{S}_\psi \cup
  \left\{
  \mathbf{X}_q|\Psi_0\rangle, \:
  \mathbf{Y}_q|\Psi_0\rangle, \:
  \mathbf{Z}_q|\Psi_0\rangle
  \right\}
  \nonumber \\
  \mathcal{S}_{\psi+2} &=&
  \mathcal{S}_\psi \cup
  \left\{
  \mathbf{X}_q|\Psi_0\rangle, \:
  \mathbf{Z}_q|\Psi_0\rangle, \:
  \mathbf{X}_q\mathbf{Z}_{q^\prime}|\Psi_0\rangle
  \right\}.
  \label{eq:psi subspaces}
\end{eqnarray}
They are orthonormal sets,
with $|\mathcal{S}_{\psi+1}| = 22$ and $|\mathcal{S}_{\psi+2}| = 64$ for the particular case of Steane's code.
Since Steane's code and the surface code are both CSS codes, they correct errors
of the form $\mathbf{X}_q \mathbf{Z}_{q^\prime}$ for all $q$ and $q^\prime$.
Checking if $|\Psi(t)\rangle$ has ventured beyond
either $\mathcal{S}_{\psi+1}$ or $\mathcal{S}_{\psi+2}$ is reasonable.

To summarize, we have three metrics
to detect a QEC circuit failure, in Eq.~(\ref{eq:failure projectors}).
For each simulation, when they are 1, the circuit has failed;
otherwise, we expect them to be 0.
\begin{eqnarray}
  P^{(L+1)}_\mathrm{fail} &=& 1 -
 \sum_{s \in L+1}
  | \langle s |\Psi(t)\rangle |^2
\nonumber \\
  P^{(\psi+2)}_\mathrm{fail} &=& 1 -
  \sum_{s \in \psi+2}
  | \langle s |\Psi(t)\rangle |^2
\nonumber \\
  P^{(\psi+1)}_\mathrm{fail} &=& 1 -
 \sum_{s \in \psi+1}
  | \langle s | \Psi(t)\rangle |^2.
  \label{eq:failure projectors}
\end{eqnarray}
Technically the sums in Eq. \eqref{eq:failure projectors} should also extend over the ancilla qubits. However, we assume that measurements are perfect, therefore when we analyze the failure metrics, we know that the ancilla will be in a product state with the data qubits, and thus the sum over them is independent and can be excluded.

Two additional metrics are useful:
the projection into $\mathcal{S}_L$,
given in Eq.~(\ref{eqn:P codeword definition}),
and the fidelity $\mathcal{F}$,
in Eq.~(\ref{eqn:fidelity definition}).
The result $\mathcal{F} = 1.0$ implies a perfect repair.
\begin{equation}
  \label{eqn:P codeword definition}
  P_\mathrm{code} = 
  \sum_{s \in \mathcal{S}_L}
  | \langle s | \Psi(t)\rangle |^2
\end{equation}
\begin{equation}
  \label{eqn:fidelity definition}
  \mathcal{F}^2 =
    \big| \langle \Psi(t) |
    \left( \alpha |0_L\rangle + \beta |1_L\rangle \right)\big|^2.
\end{equation}

Several useful bounds are shown in Eq.~(\ref{eqn:bounds}).
Physically, a low $P_\mathrm{code}$ implies a low $\mathcal{F}$,
since $|\Psi(t)\rangle$ is outside of $\mathcal{S}_L$.
The inverse does not hold, since a logical error such as
$|1_L\rangle = \mathbf{X}_L|0_L\rangle$ on an encoded $|0_L\rangle$
leaves $P_\mathrm{code} = 1$ but $\mathcal{F} = 0$.
It is also seen that $P^{(\psi+1)}_\mathrm{fail}$ is the most
stringent criteria for circuit failure.
\begin{align}
  \label{eqn:bounds}
 &  P^{(L+1)}_\mathrm{fail}  \le P^{(\psi+1)}_\mathrm{fail} \nonumber \\
 &  P^{(\psi+2)}_\mathrm{fail}  \le P^{(\psi+1)}_\mathrm{fail} \nonumber \\
 & \mathcal{F}^2  \le P_\mathrm{code} \le 1 - P^{(L+1)}_\mathrm{fail}.
\end{align}

Several computational tools to Monte Carlo estimate a
pseudo--threshold apparently restrict $|\Psi_0\rangle$
to be one of the six stabilizer states 
\begin{equation}
  \big\{ |0_L\rangle, |1_L\rangle,
  (|0_L\rangle \pm |1_L\rangle)/\sqrt{2},
  (|0_L\rangle \pm i |1_L\rangle)/\sqrt{2} \big\}.
  \label{eq:stabilizers}
\end{equation}
This makes drawing distinctions between the metrics
in Eq.~(\ref{eq:failure projectors}) problematic, as
discussed later in this paper.

\subsection{Surface Code Failure Metric}\label{sec:space_degeneracy}
The failure metrics for the Steane code was relatively straightforward since all errors map to a unique syndrome. There is a degeneracy in the surface code that complicates this somewhat. We show here how to define a failure metric analogous to that for the Steane code that takes this into account.

The logical state is generated utilizing the parity-check matrix constructed from the $\hat{X}$ stabilizers for the surface code shown later in equation \ref{eq:SCstabs}. It results in the basis states:
\begin{equation}
\begin{aligned}
\ket{0}_L = & \frac{1}{4} \Big( \ket{000000000} + \ket{110110000} + \ket{011000000} \Big.\\
& ~~~~ + \ket{101110000} + \ket{000000110} + \ket{110110110} \Big.\\
& ~~~~ + \ket{011000110} +\ket{101110110} + \ket{000011011} \Big.\\
& ~~~~+ \ket{110101011} + \ket{011011011} + \ket{101101011} \\[1ex]
& ~~~~ \Big. + \ket{000011101} + \ket{110101101} + \ket{011011101} \Big.\\
& ~~~~ + \ket{101101101} \Big)\\[2ex]
\ket{1}_L = & \; \hat{X}^{\otimes 9} \ket{0}_L
\end{aligned}
\end{equation}
that we can use to construct our logical code space as in equation \ref{eq:logical subspaces}.  Note that state $\ket{1}_L$ is the $\ket{0}_L$ state under the action of the bitwise-NOT on every binary string within it's sum.  Due to this relation, there is an equivalence in the action of single-qubit errors to both $\ket{0}_L$ and $\ket{1}_L$ with respect to the degeneracy of the single-error spaces.  For single-qubit errors, there is equivalence of the action of $\hat{X}$ errors, $\hat{X}_1 \ket{0}_L = \hat{X}_2\ket{0}_L$ and $\hat{X}_6 \ket{0}_L = \hat{X}_7 \ket{0}_L$, and also degeneracies for $\hat{Z}$ errors, $\hat{Z}_0 \ket{0}_L = \hat{Z}_3\ket{0}_L$ and $\hat{Z}_5 \ket{0}_L = \hat{Z}_8 \ket{0}_L$.  $\hat{Y}$ errors are completely non-degenerate.  We can therefore construct the failure criteria pertaining to the logical codeword space in the following manner:
\begin{widetext}
\begin{equation}
\begin{aligned}
& \mathcal{S}_L = \left\{ \ket{0}_L,\, \ket{1}_L \right\} \\
& \mathcal{S}_{L+1} = \mathcal{S}_L  \cup \left\{ \hat{X}_i \ket{0}_L,\, \hat{X}_i \ket{1}_L,\, \hat{Y}_j\ket{0}_L, \, \hat{Y}_j\ket{1}_L,\, \hat{Z}_k \ket{0}_L,\, \hat{Z}_k \ket{1}_L \right\} \; \mathrm{s.t.} \; i \neq 2,7;\;k\neq 3,8
\end{aligned}
\end{equation}
where the indices $i,j,k$ run over all data qubit indices unless otherwise specified.  Similarly, for the criteria more natural for the purpose of wavefunction simulations the spaces can be constructed as:
\begin{equation}
\begin{aligned}
& \mathcal{S}_\psi = \left\{ \ket{\Psi_0} \right\} \\
& \mathcal{S}_{\psi+1} = \mathcal{S}_\psi  \cup \left\{ \hat{X}_i \ket{\Psi_0},\, \hat{Y}_j\ket{\Psi_0}, \, \hat{Z}_k \ket{\Psi_0} \right\}\\
& \mathcal{S}_{\psi+2} = \mathcal{S}_\psi  \cup \left\{ \hat{X}_i \ket{\Psi_0}, \, \hat{Z}_k \ket{\Psi_0}, \, \hat{X}_i \hat{Z}_k \ket{\Psi_0} \right\} \; \textnormal{such  that} \; i \neq 2,7;\;k\neq 3,8\\
\end{aligned}
\end{equation}
\end{widetext}
where, again, the indices run over all qubits except for special cases.  While it appears that some single-qubit $\hat{Y}$ errors have been omitted in the space $\mathcal{S}_{\psi+2}$, the symmetry of the error space ensures that these states have been taken into account.  For instance, the state $\hat{Y}_2 \ket{\Psi_0} = i \hat{X}_2 \hat{Z}_2 \ket{\Psi_0}$ which, through the degeneracy of the $\hat{X}$ errors, can be represented by the state  $i \hat{X}_1 \hat{Z}_2 \ket{\Psi_0}$.  The assessment of the surface code will incorporate computing the overlap of output wavefunctions with the aforementioned correctable error spaces.  Because the surface code is an error \emph{correcting} code, the sucess criteria $P_{\mathrm{code}}$ and fidelity ($\mathcal{F}^2$) given in Eqs. \eqref{eqn:P codeword definition} and \eqref{eqn:fidelity definition} will be of interest as well.

\section{Steane Simulation Results}
With failure criteria now defined, we proceed to simulate the error models discussed in Sec. \ref{sec:control errors} for Steane's [[7,1,3]] code. The baseline QEC trial begins by encoding a random qubit.  The syndromes,
and the errors they detect, are given in Table~\ref{tab:syndromes}.
To ensure fault-tolerance, the three syndromes that check for $\mathbf{Z}$ errors are measured,
and then measured again.  A loop back occurs
if the three sets of syndrome bits do not match.  Otherwise,
any $\mathbf{Z}$ errors are repaired, and the circuit
repeats this for the $\mathbf{X}$ error detecting syndromes.

The syndrome measurement procedure uses Shor style ancilla
It begins by creating a cat state in the 4 ancilla qubits,
which is always error free.  The test and rejection steps seen
in Fig. 6 of \cite{aliferis2006} are thus not required.
After entanglement, the cat state is rotated 
and read out in the style discussed in Ref. \cite{divincenzo2007}.
Each ancilla is then perfectly re-initialized allowing them to be re-used for the next syndrome.

\begin{table}[h]
  \caption{\label{tab:syndromes}
    The six syndromes for Steane's [[7,1,3]] code,
    in the left column.  The right columns show the errors
    they detect.
  }
  \begin{ruledtabular}
    \begin{tabular}{c|ccccccc}
      Operator & $\mathbf{Z}_1$ & $\mathbf{Z}_2$ & $\mathbf{Z}_3$ &
      $\mathbf{Z}_4$ & $\mathbf{Z}_5$ & $\mathbf{Z}_6$ & $\mathbf{Z}_7$ \\
      \colrule
      $\mathbf{X}_2\mathbf{X}_4\mathbf{X}_5\mathbf{X}_7$ & 0 & 1 & 0 & 1 & 1 & 0 & 1 \\
      $\mathbf{X}_3\mathbf{X}_4\mathbf{X}_5\mathbf{X}_6$ & 0 & 0 & 1 & 1 & 1 & 1 & 0 \\
      $\mathbf{X}_1\mathbf{X}_4\mathbf{X}_6\mathbf{X}_7$ & 1 & 0 & 0 & 1 & 0 & 1 & 1 \\
      \colrule
      Operator & $\mathbf{X}_1$ & $\mathbf{X}_2$ & $\mathbf{X}_3$ &
      $\mathbf{X}_4$ & $\mathbf{X}_5$ & $\mathbf{X}_6$ & $\mathbf{X}_7$ \\
      \colrule
      $\mathbf{Z}_2\mathbf{Z}_3\mathbf{Z}_6\mathbf{Z}_7$ & 0 & 1 & 1 & 0 & 0 & 1 & 1 \\
      $\mathbf{Z}_1\mathbf{Z}_3\mathbf{Z}_5\mathbf{Z}_7$ & 1 & 0 & 1 & 0 & 1 & 0 & 1 \\
      $\mathbf{Z}_1\mathbf{Z}_2\mathbf{Z}_3\mathbf{Z}_4$ & 1 & 1 & 1 & 1 & 0 & 0 & 0 \\
    \end{tabular}
  \end{ruledtabular}
\end{table}

\subsection{Pauli Error Model}\label{sec:pauli_error_model}

\begin{figure}[h]
  \includegraphics[width=8.4cm]{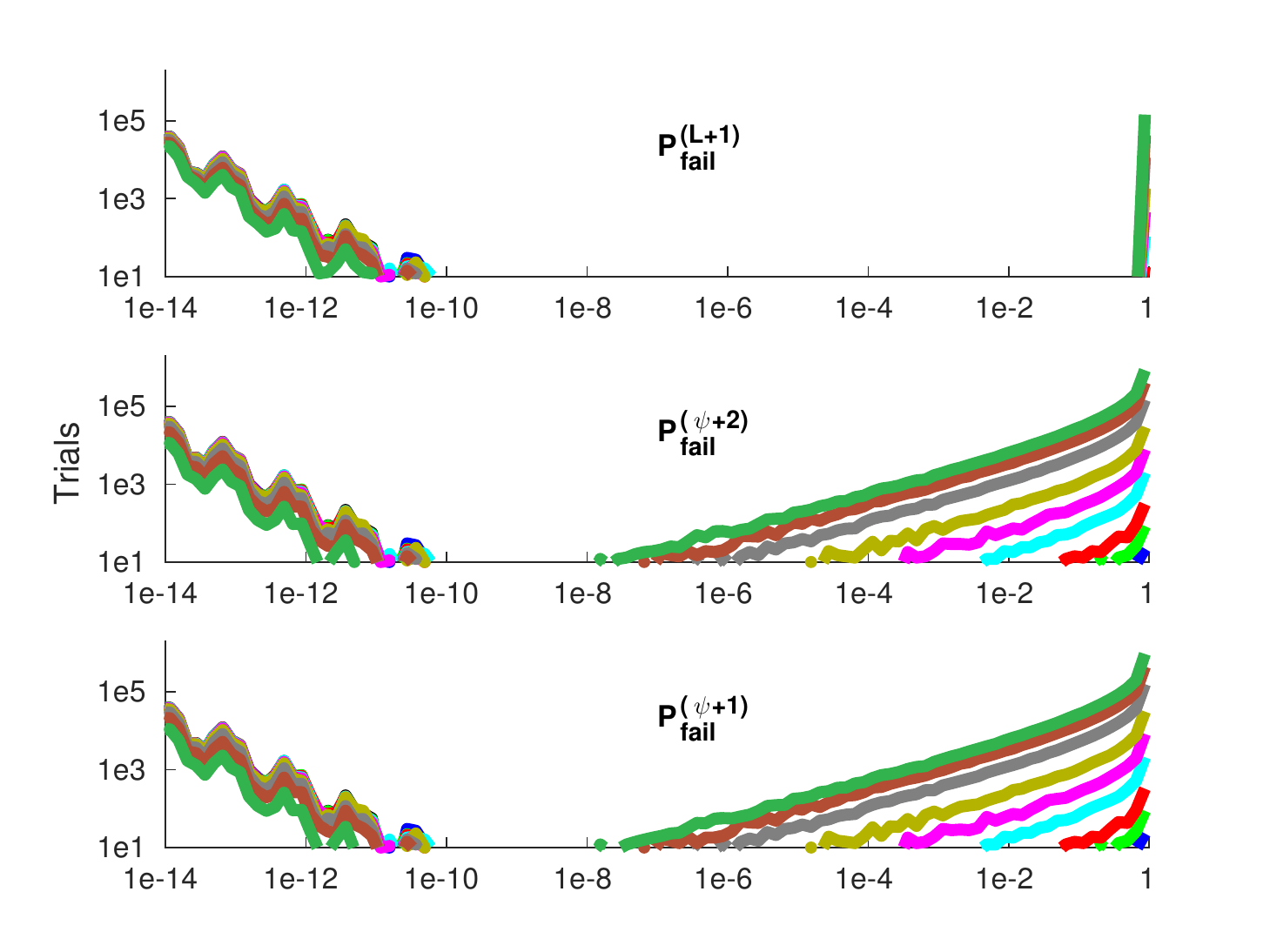}
  \caption{\label{fig:pfail pauli histograms}
    Histograms of the Steane QEC failure metrics,
    from $3\times 10^6$ trials of the Steane QEC circuit,
    using the Pauli error model
    with $p = .00005, .0001, .0002, .0005, .001, .002, .005, .01, .02$.
    $P^{(L+1)}_\mathrm{fail}$ (top) is always binomially distributed,
    but $P^{(\psi+2)}_\mathrm{fail}$ (middle) and $P^{(\psi+1)}_\mathrm{fail}$ (bottom)
    never are.
  }
\end{figure}

\begin{figure}[h]
  \includegraphics[width=8.4cm]{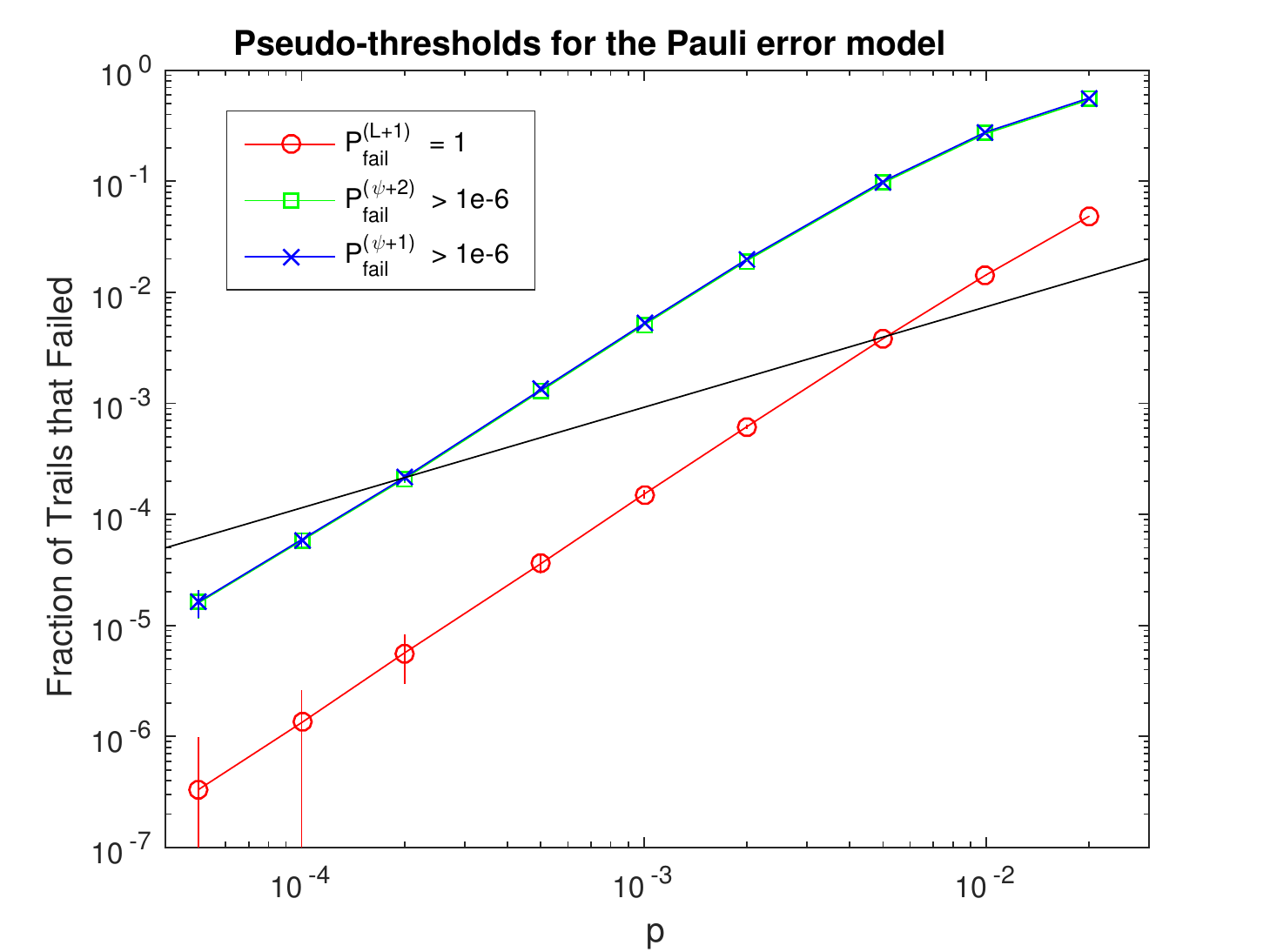}
  \caption{\label{fig:pfail pauli}
    The Monte--Carlo simulation of the failure rate for
    the Steane QEC circuit, using the Pauli error model.
    The red circles are the fraction of trials with $P^{(L+1)}_\mathrm{fail} = 1$.
    The blue and green symbols mark estimates of failure,
    based on the criteria of $P^{(\psi+2)}_\mathrm{fail} > 10^{-6}$
    and $P^{(\psi+1)}_\mathrm{fail} > 10^{-6}$.
    The vertical bars represent the 95\% confidence level in the estimates.
    The black line marks the pseudo-threshold criteria of $P_\mathrm{fail}(p) = p$.
  }
\end{figure}

The Steane QEC circuit was run $3\times 10^6$ times
at nine different $p$ values.  Histograms of each failure criteria
are shown in Fig.~\ref{fig:pfail pauli histograms}.
The metric $P^{(L+1)}_\mathrm{fail}(p)$ is binominally distributed (within numerical accuracy),
and has the obvious interpretation.
The fraction of trials with $P^{(L+1)}_\mathrm{fail}(p) = 1$ for a given $p$
is shown by the red curve in Fig.~\ref{fig:pfail pauli}.
A pseudo-threshold of $\approx 0.005$ can be observed.
The curve appears similar to examples provided in \cite{svore2006},
and gives confidence that the wavefunction simulations can
reproduce the standard pseudo-threshold.

The difficulty with $P^{(\psi+1)}_\mathrm{fail}$ and $P^{(\psi+2)}_\mathrm{fail}$
is that the spaces of Eq.~(\ref{eq:psi subspaces}) are not aligned
with the 6 stabilizer states of Eq.~(\ref{eq:stabilizers}).
When the code is run with the $|\Psi_0\rangle$ restricted to be
from the set of Eq.~(\ref{eq:stabilizers}), then all three metrics in
Eq.~(\ref{eq:failure projectors}) are binomially distributed. We note here that all Gottesman-Knill and error propagation simulations either require or implicitly assume that input states are Stabilizer states. Even for Pauli errors, this restriction already changes the output failure distribution away from what will actually happen when arbitrarily encoded states must be protected.
Since, arbitrary encoded states cannot be abandoned, so how can this result
be understood?

\begin{figure}[h]
  \includegraphics[width=8.4cm]{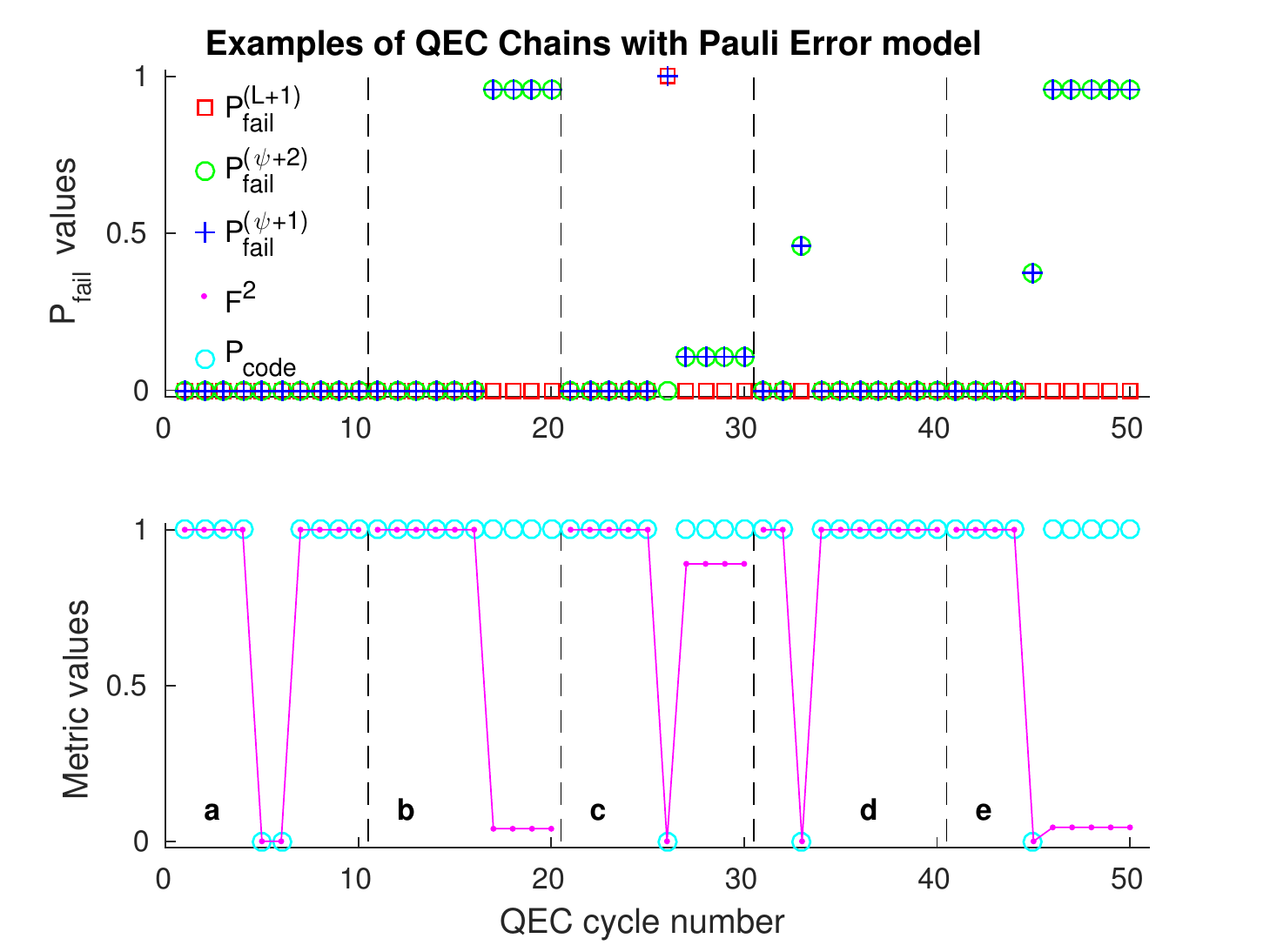}
  \caption{\label{fig:pauli chain examples}
    Shown are five different runs of 10 consecutive cycles of
    Steane's [[7,1,3]] QEC circuit, using the Pauli error model.
    At top are plots of the three metrics for QEC failure under consideration.
    At bottom are plots of $\mathcal{F}^2$ and $P_\mathrm{code}$.
  }
\end{figure}

First ask: does {\it every} non-zero value of
$P^{(\psi+1)}_\mathrm{fail}$ or $P^{(\psi+2)}_\mathrm{fail}$ indicate a circuit failure?
One way to examine this is to run chains of QEC circuits, where the input of next QEC cycle is the output of the previous one. In such simulations, $P_\mathrm{code}$ and $\mathcal{F}$ transiently drop from 1.0, but then they recover, as seen in Fig.~\ref{fig:pauli chain examples} (a). This is a case where weight-1 errors escape a QEC cycle, only to be repaired in the next cycle.  By chance, this can occur sequentially, but the $P_\mathrm{fail}$ metrics remain 0.0, indicating the errors were not fatal ones.

In contrast, if $\mathcal{F}^2 < 1$ and $P_\mathrm{code} = 1$, then the encoded qubit was corrupted, and clearly the QEC circuit failed. These situations are seen in Fig.~\ref{fig:pauli chain examples} (b) to (e). Scatterplots of large samples with $p = 0.02$ did not find any other distinct behaviors, so we consider how to explain each case.

\subsubsection{Case b}
 Within a single cycle, $P_\mathrm{code}=1$ and $\mathcal{F}<1$, indicating a corrupted but encoded qubit. There are many insertions of 3 Pauli errors that caused this.  An example is $\mathbf{X}_4$, $\mathbf{X}_5$, and then $\mathbf{X}_6$, all during the measurement of the $\mathbf{Z}$ error detecting syndromes.  Within Steane's code, $\mathbf{X}_5\mathbf{X}_4 \equiv \mathbf{X}_6\mathbf{X}_L$, so the $\mathbf{X}$ repair returns $\mathbf{X}_L|\Psi_0\rangle$, a properly encoded but corrupted state.

 In the failed QEC cycle, it is found that $P^{(L+1)}_\mathrm{fail} = 0$,
 yet $P^{(\psi+2)}_\mathrm{fail} = P^{(\psi+1)}_\mathrm{fail} = 0.9597$ and $\mathcal{F} = 0.2008$.
 The explanation comes by considering the starting state of Eq.~(\ref{eq:starting state}).
 For a $\mathbf{X}_L$ error, $P^{(\psi+1)}_\mathrm{fail} = 1 - \sin^4 \theta \, \cos^2 \phi$.
 If the success criteria is set by demanding a threshold,
 say $P^{(\psi+1)}_\mathrm{fail} > 10^{-6}$, then irreparable errors
 are allowed when encoded states are ``lucky'' enough to be nearly an
 eigenstate of that error.  For a threshold of $10^{-6}$,
 errors are allowed if they shift $|\Psi_0\rangle$ by $< 0.05^\circ$
 on the Bloch sphere.  Whether this suffices will depend on
 the needs of the quantum algorithm.

\subsubsection{Case c} 
In this trial, two fault locations had errors.
Referring to Table~\ref{tab:syndromes}, the $\mathbf{Z}$ error detection syndromes finished without errors, but a $\mathbf{Z}_6$ error occurs after the first two $\mathbf{X}$ syndrome measurements. Since these are checking for $\mathbf{X}$ errors, the syndrome is (0,0,0). In the second round, a $\mathbf{X}_7$ error occurs before the last syndrome.  This error should return (1,1,0), but we are on the last syndrome, so (0,0,0) results. As the consistency check is passed, no errors are detected. For this $p^2$ process, an error of $\mathbf{X}_7\mathbf{Z}_6$ escapes.  These combination of $\mathbf{X}_q\mathbf{Z}_{q^\prime}$ for $q \ne q^\prime$ are the only errors that distinguish $P^{(\psi+1)}_\mathrm{fail}$ from $P^{(\psi+2)}_\mathrm{fail}$. In the spirit of Ref. \cite{aliferis2006}, the more stringent weight-1 error criteria of $P^{(\psi+1)}_\mathrm{fail} > 10^{-6}$ is favored.

\subsubsection{Cases d and e} 
This appears to show cases where $P^{(\psi+1)}_\mathrm{fail} > 10^{-6}$
cannot distinguish all errors.  Both (d) and (e) would be detected as errors,
yet the permanent loss of $\mathcal{F}$ in (d) is not observed in (e).
An examination of several QEC chain trials suggests the answer.

In one case, during the second measurement of the syndromes to detect $\mathbf{X}$ errors, but just before the consistency check, the following Pauli operators were inserted: S, $\mathbf{X}_5$, S, $\mathbf{X}_6$, S, where $S$ implies a syndrome measurement. The first syndrome returns 0.  An $\mathbf{X}_5$ error should give (1,0,0), so the second syndrome returns 0. At the next error, $\mathbf{X}_6\mathbf{X}_5 \equiv \mathbf{X}_4\mathbf{X}_L$, and the syndrome for an $\mathbf{X}_4$ error is (0,1,0). Thus, the last measurement yields (0,0,0), consistent with the first syndrome measurments. This $p^2$ process allowed a corrupted qubit with a weight-1 error to pass.

In another case, in the second measurement of the $\mathbf{X}$ error detecting syndromes, this sequence was found: S, $\mathbf{X}_3$, S, $\mathbf{X}_2$, S. As before, this dances around the detection table, and $\mathbf{X}_6\mathbf{X}_L$ escapes. In the next QEC cycle, however, a single $\mathbf{X}_3$ error occurs within the $\mathbf{Z}$ detecting syndromes. Only a $\mathbf{X}_2$ error is left for the QEC circuit to repair. By random chance, a $p$ process has fixed an unrepairable $p^2$ error.

Three cases involving $p^2$ processes with outcomes like in (e) were examined.  They all had the same result: by random chance, the next QEC cycle had a $p$ process that converted the weight-2 error into a weight-1 error. As this cannot be assigned to the role of the QEC circuit, we favor the criteria $P^{(\psi+1)}_\mathrm{fail} > 10^{-6}$ to be a good detector of QEC circuit failure.

\subsection{Pulse-Area Error Model}\label{sec:unitary_errors}

We now use the pulse area error model outlined in Eq.~(\ref{eq:sigma definition}).
The reader should keep in mind the following argument in support of a stochastic error model.
Consider the QEC circuit as an operator expansion:
\begin{equation}
  \cdots \mathbf{U}_5 \mathbf{U}_4 \mathbf{\Omega}_3 \mathbf{U}_2 \mathbf{U}_1 |\Psi_0\rangle
\end{equation}
There are measurements $\mathbf{\Omega}$ and unitary gates $\mathbf{U}$.
Following Eq.~(\ref{eq:sigma definition}), and using
$\mathbf{U}_n^2 = \mathbf{1}$, this becomes
\begin{widetext}
\begin{equation}
  \cdots
  \mathbf{\Omega}_3
  \left( \cos\frac{\pi\sigma r_2}{2}\,\mathbf{U}_2 - i\sin\frac{\pi\sigma r_2}{2}\,\mathbf{1} \right)
  \left( \cos\frac{\pi\sigma r_1}{2}\,\mathbf{U}_1 - i\sin\frac{\pi\sigma r_1}{2}\,\mathbf{1} \right)
  |\Psi_0\rangle
  \label{eqn:operator expansion}
\end{equation}
\end{widetext}
Now suppose $\sigma \ll 1$.  In that case, the linear terms
in $\sigma$ simply have each gate $\mathbf{U}_n \rightarrow \mathbf{1}$.
This superficially resembles the Pauli error model:  insert a Pauli operator,
which cancels the gate, with amplitude $\sigma r_n$ at gate $n$.
The {\it sum} of these terms in Eq.~(\ref{eqn:operator expansion})
creates $|\Psi(t)\rangle$.  Using the argument in support of the stochastic error model, each syndrome
measurement $\mathbf{\Omega}$ selects one of these terms at a time, so the incoherent
sum only needs to be considered.  That is very useful,
because even if $\sigma^2 \ll \sigma$, with the 144 gates of the Steane syndrome extraction circuitry,
there are 10,296 dual gate $\mathcal{O}(\sigma^2)$ terms to add up,
which would be difficult.  Yet, in order to fully select one error,
6 syndromes must be measured, even while errors continue to accumulate.
The numerical simulation can be used to check this approximation.

\begin{figure}[h]
  \includegraphics[width=8.4cm]{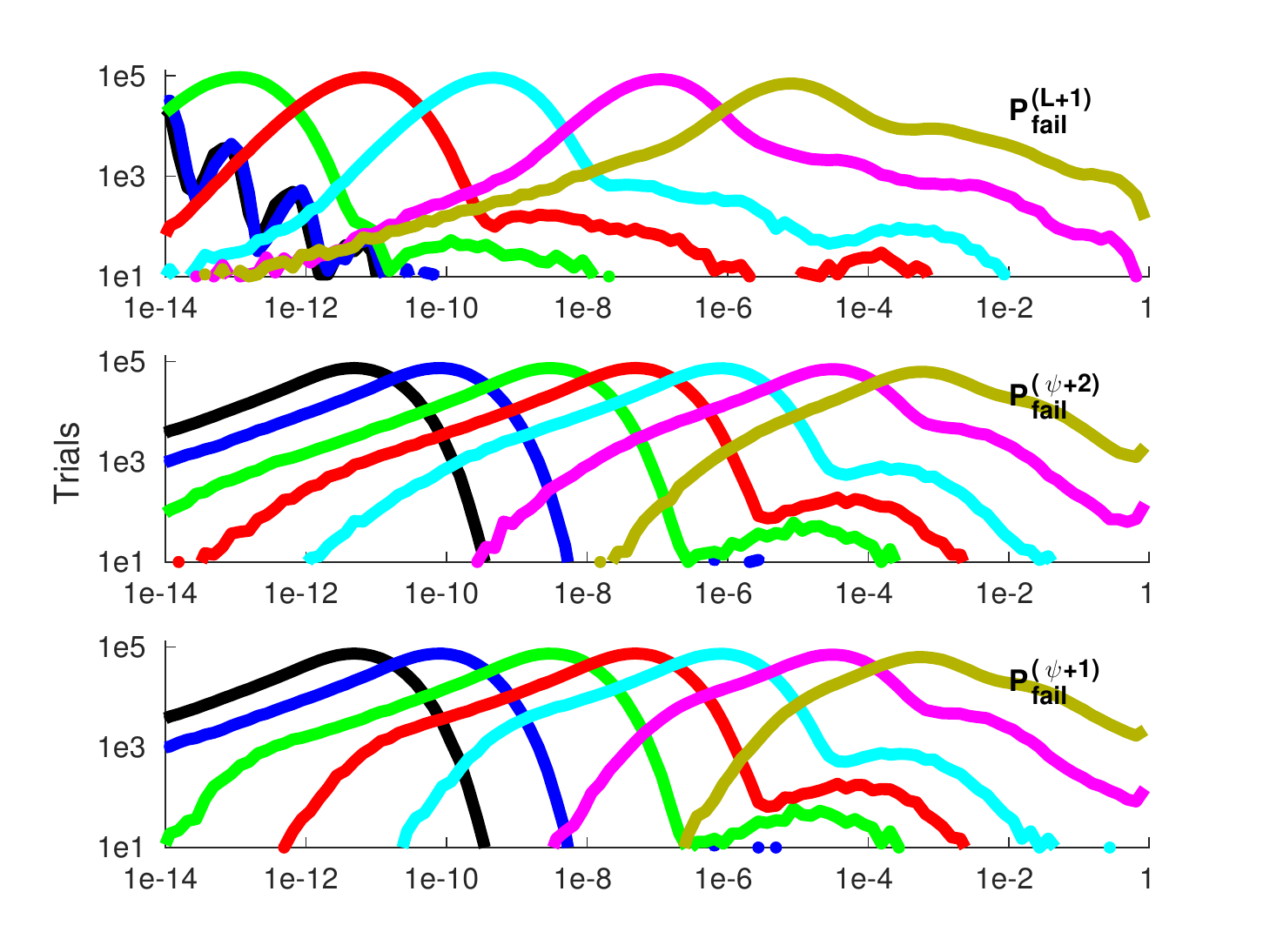}
  \caption{\label{fig:control pfail histograms}
Histograms of the QEC circuit failure metrics, for $1\times 10^6$ trials of $\sigma$ = .001, .002, .005, .010, .020, .050, and .100. The curves proceed from left to right on the bottom panel.
  }
\end{figure}

\begin{figure}[h]
  \includegraphics[width=8.4cm]{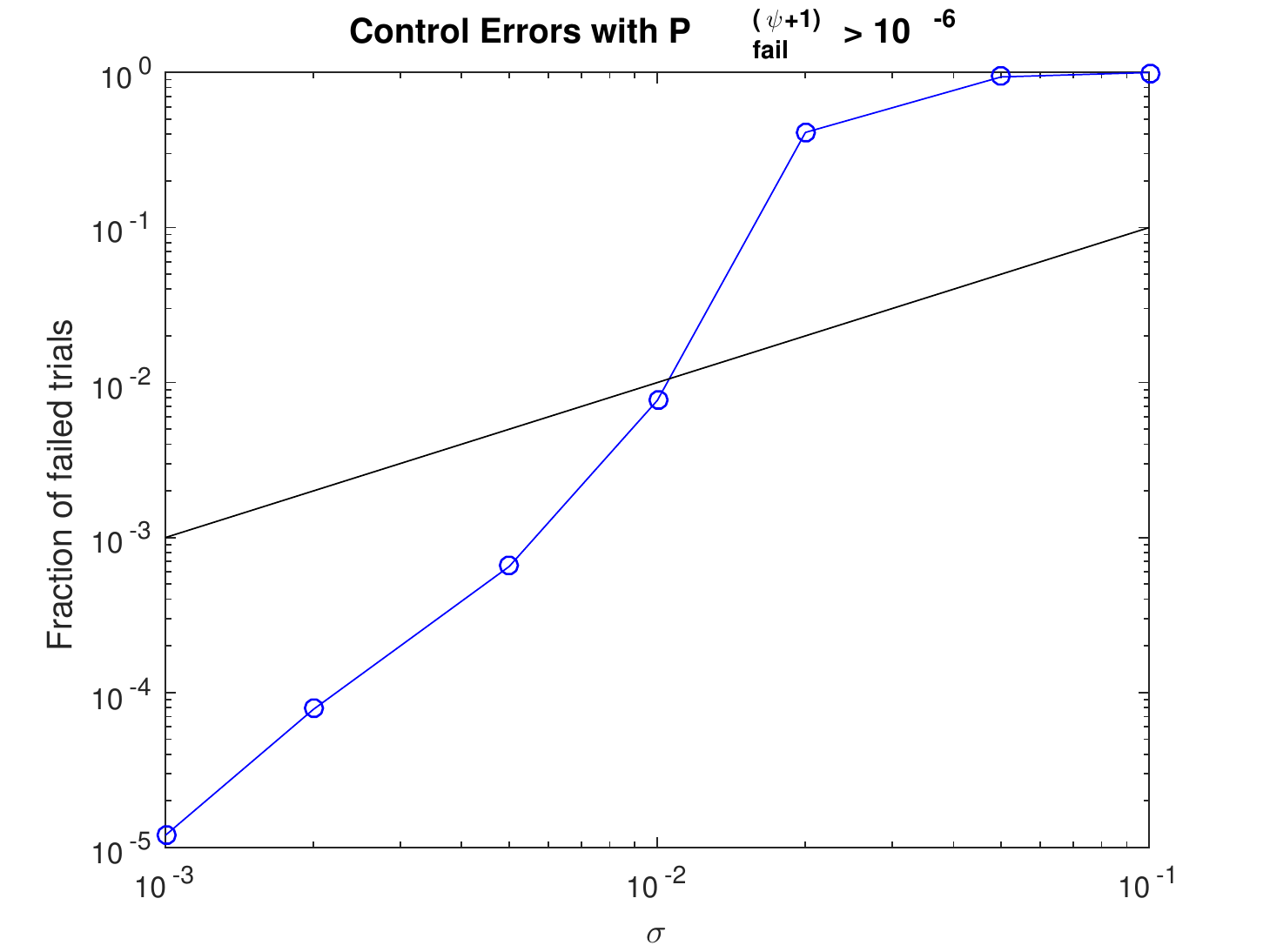}
  \caption{\label{fig:control pseudothreshold}
Using the criteria $P^{(\psi+1)}_\mathrm{fail} > 10^{-6}$ on the $10^6$ trials of the Steane QEC circuit, under the pulse-area error model, gives rise to this failure rate curve.  The failure rate does not follow the expected linear slope. The black line is the unit line.
  }
\end{figure}

To see how the Steane code handles these pulse area errors,
$10^6$ trials with $0.001 \le \sigma \le 0.100$ are shown in
Fig.~\ref{fig:control pfail histograms}.  All three failure metrics
are shown for completeness.
First, note that the distributions are {\it very} broad.
A curve fit to the distribution of
$P^{(\psi+1)}_\mathrm{fail}(\sigma = .01)$ showed that
roughly 75\% of the trials could be described by
the log-normal distribution of the form $\exp(-a \log^2(b P_\mathrm{fail}))$.
The maximum likelihood scales as $\mathrm{ML}(P^{(\psi+1)}_\mathrm{fail}) \approx 5 \sigma^4$.
Employing the $P^{(\psi+1)}_\mathrm{fail} > 10^{-6}$ critera for failure results
in a failure-versus-$\sigma$ curve in Fig.~\ref{fig:control pseudothreshold}.
For small $\sigma$, it varies as $\approx 200 \sigma^{2.5}$,
but comparison with Fig.~\ref{fig:pfail pauli} reveals a much more sigmodial shape.

\begin{figure}[h]
  \includegraphics[width=8.4cm]{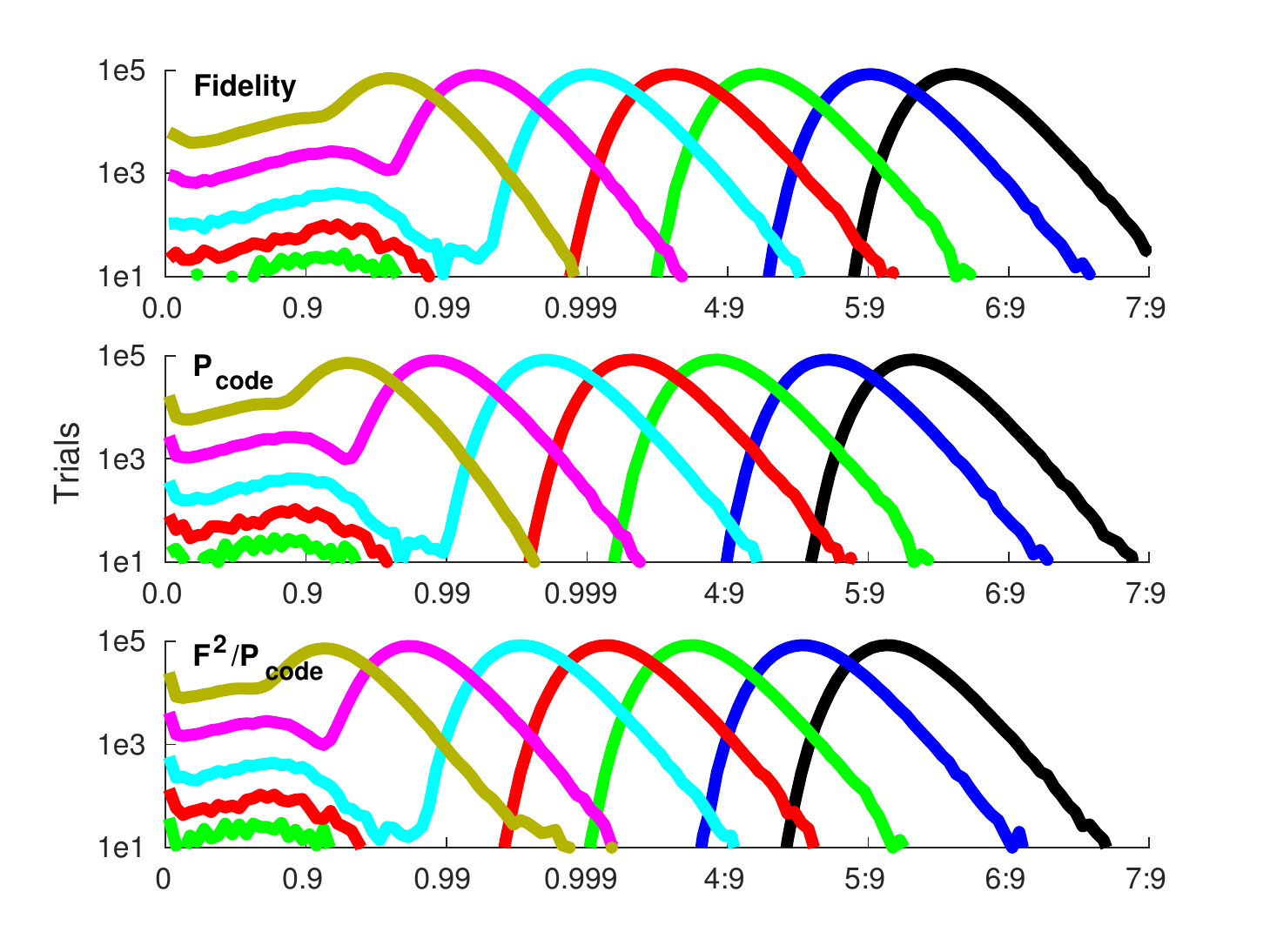}
  \caption{\label{fig:control fidelity}
The success metrics $\mathcal{F}$, $P_\mathrm{code}$, and the ratio, for the pulse-area error model, acting on the Steane [[7,1,3]] QEC circuit. Values of $\sigma$ correspond to those of Fig.~\ref{fig:control pfail histograms}.
  }
\end{figure}

In the Pauli error model, a successful QEC cycle competely restores the encoded qubit. What about the pulse area error model? Fig.~\ref{fig:control fidelity} shows log-log plots for the distributions of $\mathcal{F}$ and $P_\mathrm{code}$. Perfect outcomes ($\mathcal{F} = P_\mathrm{code} = 1$) are not apparent, but it may be that the part of $|\Psi(t)\rangle$ outside of $\mathcal{S}_L$ is still repairable.  What about the part inside? Consider the metric in Eq.~(\ref{eqn:ratio}).
\begin{equation}
  \mathcal{F}^2/P_\mathrm{code} =
  \frac{\displaystyle |\langle \Psi_0|\Psi(t)\rangle |^2}
       {\displaystyle |\langle 0_L|\Psi(t)\rangle |^2 + |\langle 1_L|\Psi(t)\rangle|^2}
       \label{eqn:ratio}
\end{equation}
It normalizes the part of $|\Psi(t)\rangle$ still in $\mathcal{S}_L$.  If it is $< 1.0$, then $|\Psi(t)\rangle$ has rotated from $|\Psi_0\rangle$ within the codeword space. This is an unrepairable error. Histograms of this metric are shown at the bottom of Fig.~\ref{fig:control fidelity}. For $\sigma = 0.005$, the most likely value of $\mathcal{F}^2/P_\mathrm{code} \approx 0.0001$. Thus, a quantum calculation that encodes digital numbers into amplitudes will only be accurate to approximately 4 digits.  This distribution scales like $4.6 \sigma^4$, so to achieve float64 precision (16 decimal digits) requires $\sigma \approx 0.00007$.  This is analogous to round-off errors in finite precision classical computers. We note that this error is likely to be reduced as code distance is increased, but lacking a distance 5 simulation we cannot show this.

We finally note that, like the Pauli error model,
we have simulated strings of QEC circuits to examine
the long term behavior (not shown).  This small errors
noted here did not grow over multiple cycles, but remained bounded.
The QEC circuit is working, but the behavior is much more
complicated than the Pauli error model suggests.

\section{Surface Code Simulation Results}\label{sec:surface code}
We have demonstrated that the failure distributions for the pulse area error model are qualitatively different than those generated by purely Pauli errors. For completeness we run a similar analysis on a surface code simulation. The particular variant we choose is the tilted-17 surface code, which is a variation on the distance 3 surface code that requires less data and ancilla qubits compared to conventional distance 3 surface code \cite{OptimalResourcesTop, SCLattSurgery,FTQCAnyons}.  A diagram of the qubit and stabilizer layout of the code is shown in Fig. \ref{fig:surfacecode}.  A list of the eight stabilzers are shown below:
\begin{equation}
\begin{tabular}{ccc}
$\hat{X}_0 \hat{X}_1 \hat{X}_3 \hat{X}_4$ & & $\hat{Z}_0 \hat{Z}_3$ \\
$\hat{X}_1 \hat{X}_2$ & & $\hat{Z}_1 \hat{Z}_2 \hat{Z}_4 \hat{Z}_5$ \\
$\hat{X}_6 \hat{X}_7$ & & $\hat{Z}_3 \hat{Z}_4 \hat{Z}_6 \hat{Z}_7$ \\
$\hat{X}_4 \hat{X}_5 \hat{X}_7 \hat{X}_8$ & & $\hat{Z}_5 \hat{Z}_8$ \\
\end{tabular}
\label{eq:SCstabs}
\end{equation}
The measurement of the weight-4 stabilizers are scheduled in such a way that single qubit errors on the ancilla qubits propagate to, at most, two-qubit errors on the data in a manner such that the propagation direction is perpendicular to the direction of the logical operator of the surface code resulting in fault-tolerance \cite{LowDSCRealNoise}. An additional requirement for fault-tolerance is the repetitive measurement (typically for $d$ measurement rounds) of the stabilizers before applying a correction operation.

\begin{figure}[t]
\centering
\includegraphics[width=0.5\textwidth]{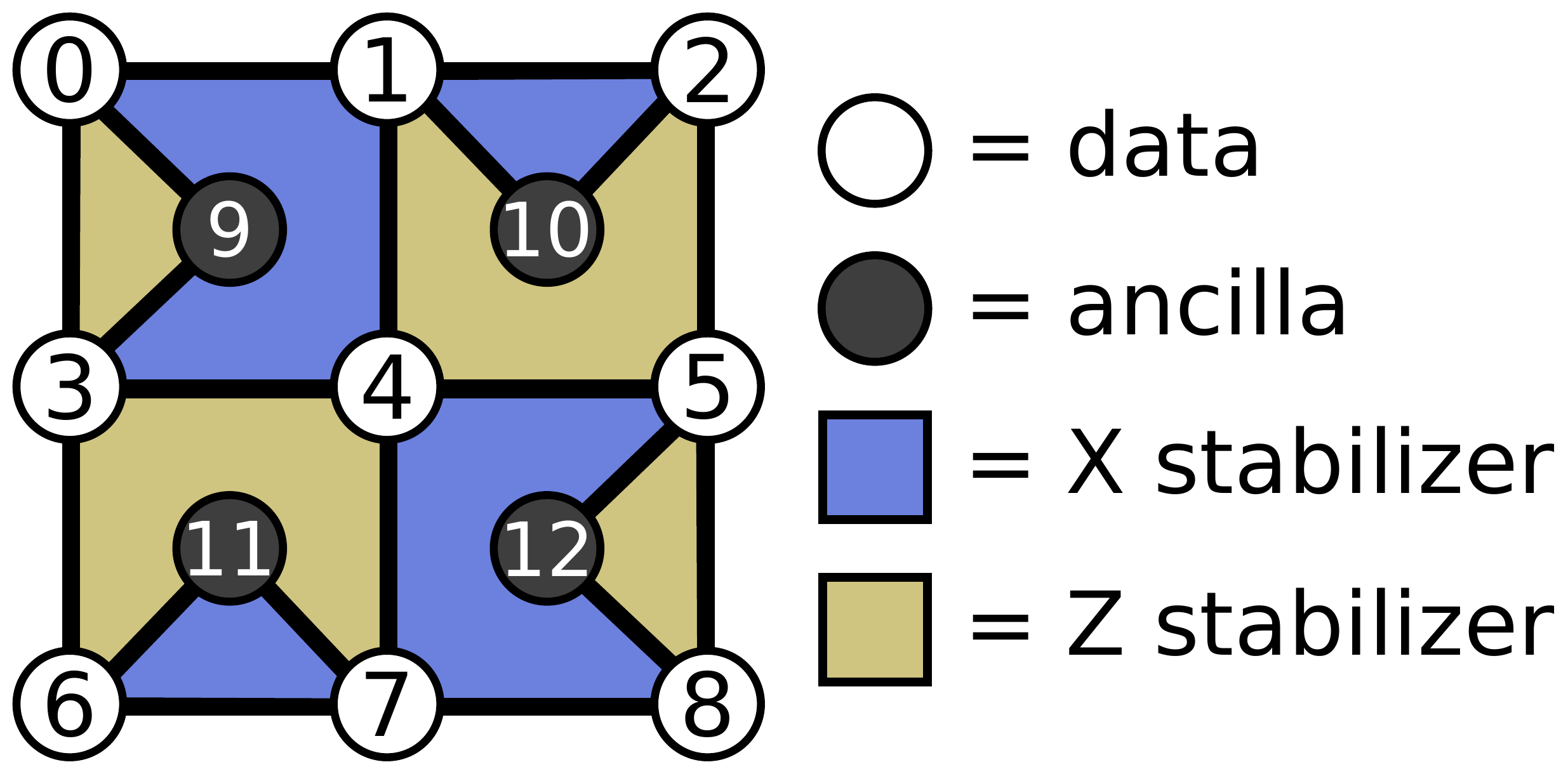}
\caption{\label{fig:surfacecode}The tilted-17 surface code (distance 3).  A $3\times3$ layout of data qubits and a set of weight 2 and 4 stabilizers that, with an appropriate circuit schedule, is a fault-tolerant implementation of an error correcting code.}
\end{figure} 

For this study, a lookup table decoder was implemented for error correction \cite{LowDSCRealNoise}.  This method utilizes a small set of syndrome processing rules that is equivalent to applying a minimum weight perfect matching algorithm \cite{MWPMSC,EdmondsOrig,EdmondsAlgo} to only nearest neighbor syndrome pairs \cite{LowDSCRealNoise} which, for a distance 3 code, is sufficient for error correction. There is some freedom to how one can schedule the syndrome extraction routine. For this manuscript we use a ``single-shot'' detection and correction cycle where we perform $\hat{X}$ followed by $\hat{Z}$ stabilizer measurements, repeat them three times to ensure fault tolerance, then decoding/correction is performed once. 

\subsection{Pauli Error Model} \label{ssec:SCpauli}

\begin{figure}[h]
\includegraphics[clip,width=8.4cm,trim= 4cm 9.1cm 4cm 9.1cm]{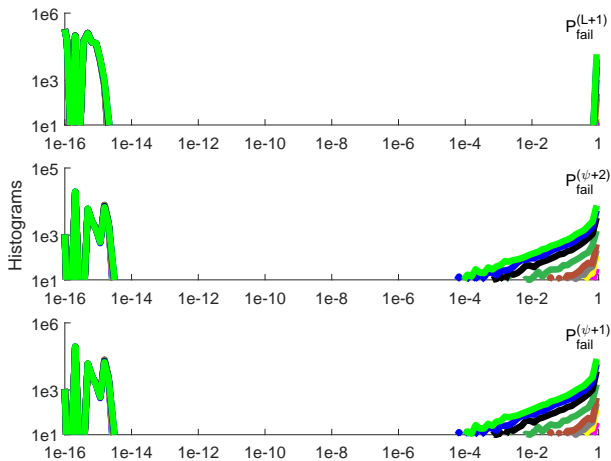}
  \caption{\label{fig:SC_pauli_histograms}
    Histograms of the failure criteria for the tilted surface code with the Pauli error model. $1 \times 10^6$ samples were accumulated per error rate.  Pauli error rates: $p$ =  0.000001, 0.0000025, 0.000005, 0.0000075, 0.00001, 0.000025, 0.00005, 0.000075, 0.0001, 0.0002, 0.0004, 0.0006, 0.0008 generate the curves from left to right respectively.
  }
\end{figure}

The ``single-shot'' error correction routines were each run $1 \times 10^6$ times at each error rate $p$ =  0.000001, 0.0000025, 0.000005, 0.0000075, 0.00001, 0.000025, 0.00005, 0.000075, 0.0001, 0.0002, 0.0004, 0.0006, 0.0008.  We show the failure histograms generated from the surface code simulations, in Fig. \ref{fig:SC_pauli_histograms}. Besides slightly shifted values for the failure histograms there is no qualitative difference between the surface code results compared to the Steane code shown in Fig. \ref{fig:pfail pauli histograms}. This is not a surprising or unexpected result.

We observe a pseudo-threshold of $\sim 3 \times 10^{-5}$ from these surface code simulations.  Note that our reported pseudo-thresholds appear to be below the value reported in \cite{LowDSCRealNoise} of $3 \times 10^{-4}$.  The simulations in \cite{LowDSCRealNoise} implement $\ket{0}_L$ as in input wavefunction while our simulations randomly sample a vector in the logical code word space $1/\sqrt{2} \left(\alpha \ket{0}_L + \beta \ket{1}_L \right)$.  Also, there is a discrepancy between the labeling of the error rates between this study and the study in \cite{LowDSCRealNoise} where our recorded values of $p$ is equivalent to $p/3$ in \cite{LowDSCRealNoise}.  By fixing our initial state to just $\ket{0}_L$ and using the $P_{\mathrm{fail}}^{\left(L+1 \right)}$ criteria, we obtain a pseudo-threshold of $\approx 4 \times 10^{-5}$ which, in the language of \cite{LowDSCRealNoise}, is reported as $p_{th} \approx 1.2 \times 10^{-4}$; a comparable value.

\begin{figure}[h]
\includegraphics[clip,width=8.4cm,trim= 4cm 9cm 4cm 9.1cm]{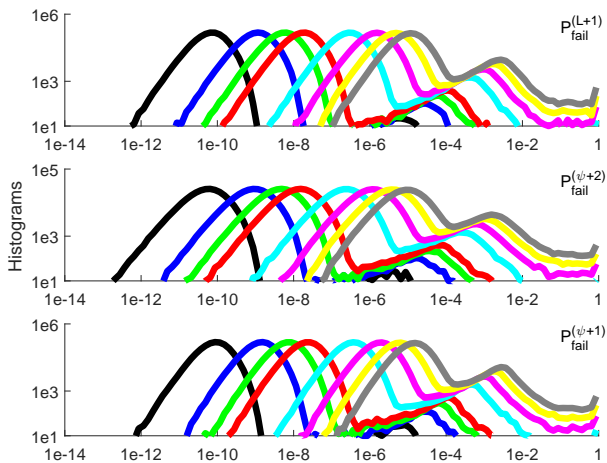}
  \caption{\label{fig:SC_pulse_histograms}
    Histograms of the failure criteria for the tilted surface code with the pulse area error model. $1 \times 10^6$ samples were accumulated per error rate.  Pulse area error strengths: $\sigma$ = 0.0025, 0.005, 0.0075, 0.01, 0.02, 0.03, 0.04, 0.05. generate the curves from left to right.
  }
\end{figure}

\subsection{Pulse-Area Error Model}

To compare the failure criteria of the surface code between the Pauli error model and pulse-area error model, we perform surface code simulations using the exact same circuit as for the Pauli error model just shown. We use the following error strengths, $\sigma$: 0.0025, 0.005, 0.0075, 0.01, 0.02, 0.03, 0.04, 0.05 ($1\times10^6$ samples per error rate).  Broad, heavy-tailed distributions for the failure criteria are again observed as shown in Fig. $\ref{fig:SC_pulse_histograms}$ and appear to follow a log-normal distribution in a majority of the cases. Once again, just as the failure metrics for the Pauli error model and pulse area error model varied drastically with the Steane code, the surface code displays similar behavior. There is no simple map between the Pauli error model and the pulse area error model, with the pulse-area error model leading to quite broad and heavy-tailed distributions.
\section{Conclusions}
We have presented a wavefunction simulation of Steane's [[7,1,3]] quantum error correcting code and the tilted-17 surface code for both the stochastic Pauli error model and the coherent pulse-area error model for randomly encoded states.  The usual failure criteria had to be modified in order to account for the different frame that non-stabilizer states occupy. A comparison of Fig.~\ref{fig:pfail pauli histograms} and Fig.~\ref{fig:control pfail histograms} shows that the two error models result in markedly different failure distributions.  Pauli errors tend to draw $P^{\psi+1}_\mathrm{fail}(p)$ down from 1.0 with increasing $p$.  These tails empirically vary as $\sqrt{P^{\psi+1}_\mathrm{fail}}$.  In contrast, for increasing $\sigma$ in the pulse-area error model, the distribution rises from small $P^{\psi+1}_\mathrm{fail}$, with the most likely values scaling as $\sigma^4$. We conjecture that the heavy tails in the pulse-area error model distribution are an indicator of the strongly negative impact that purely coherent errors can have on QEC \cite{sander2016, ExpFTThresh}.

These results also demonstrate that any result attempting to approximate arbitrary physical error channels with stochastic Pauli error channels must do so only when modeling certain system-bath type models. We have shown that one cannot rely on the assumption that syndrome measurements cut off the interfering pathways. A stochastic error model is only appropriate when the source of noise is due only to errors that are entangled with bath states that are mutually orthogonal to all other bath states. Of course, this is not true in general with coherent errors being the extreme counter-example. A simulation employing such an approximation, while it may be able to bound or approximate a mean threshold, will not be able to reproduce the output distribution that contains information from the fully coherent sum of all fault paths. As we have shown here, even for very small perturbations, the coherent addition of these fault paths cause marked differences in the output statistics. This also implies that any analysis that contains a perfect QEC measurement cycle and correction will, by design, cut off these coherent pathways, also corrupting the output statistics.

Finally, we note that these output distributions can be used as a more robust metric for QEC performance than just considering the logical error rate. They would allow one to examine whether ideas to randomize coherent errors, such as randomized compiling \cite{PhysRevA.94.052325} do indeed help by making the errors look more ``Pauli'' like to QEC. The authors do not expect to see any benefit from randomized compiling as the evolution for a single trajectory would still be completely coherent, but this is a question for further investigation.

\acknowledgments{We acknowledge useful conversations with Joan Hoffmann and Ken Brown. This project was supported by the Intelligence Advanced Research Projects Activity via Department of Interior National Business Center contract number 2012-12050800010. The U.S. Government is authorized to reproduce and distribute reprints for Governmental purposes notwithstanding any copyright annotation thereon. The views and conclusions contained herein are those of the authors and should not be interpreted as necessarily representing the official policies or endorsements, either expressed or implied, of IARPA, DoI/NBC, or the U.S. Government.}

\providecommand{\noopsort}[1]{}\providecommand{\singleletter}[1]{#1}%


\begin{thebibliography}{45}%
\makeatletter
\providecommand \@ifxundefined [1]{%
 \@ifx{#1\undefined}
}%
\providecommand \@ifnum [1]{%
 \ifnum #1\expandafter \@firstoftwo
 \else \expandafter \@secondoftwo
 \fi
}%
\providecommand \@ifx [1]{%
 \ifx #1\expandafter \@firstoftwo
 \else \expandafter \@secondoftwo
 \fi
}%
\providecommand \natexlab [1]{#1}%
\providecommand \enquote  [1]{``#1''}%
\providecommand \bibnamefont  [1]{#1}%
\providecommand \bibfnamefont [1]{#1}%
\providecommand \citenamefont [1]{#1}%
\providecommand \href@noop [0]{\@secondoftwo}%
\providecommand \href [0]{\begingroup \@sanitize@url \@href}%
\providecommand \@href[1]{\@@startlink{#1}\@@href}%
\providecommand \@@href[1]{\endgroup#1\@@endlink}%
\providecommand \@sanitize@url [0]{\catcode `\\12\catcode `\$12\catcode
  `\&12\catcode `\#12\catcode `\^12\catcode `\_12\catcode `\%12\relax}%
\providecommand \@@startlink[1]{}%
\providecommand \@@endlink[0]{}%
\providecommand \url  [0]{\begingroup\@sanitize@url \@url }%
\providecommand \@url [1]{\endgroup\@href {#1}{\urlprefix }}%
\providecommand \urlprefix  [0]{URL }%
\providecommand \Eprint [0]{\href }%
\providecommand \doibase [0]{http://dx.doi.org/}%
\providecommand \selectlanguage [0]{\@gobble}%
\providecommand \bibinfo  [0]{\@secondoftwo}%
\providecommand \bibfield  [0]{\@secondoftwo}%
\providecommand \translation [1]{[#1]}%
\providecommand \BibitemOpen [0]{}%
\providecommand \bibitemStop [0]{}%
\providecommand \bibitemNoStop [0]{.\EOS\space}%
\providecommand \EOS [0]{\spacefactor3000\relax}%
\providecommand \BibitemShut  [1]{\csname bibitem#1\endcsname}%
\let\auto@bib@innerbib\@empty
\bibitem [{\citenamefont {Shor}(1995)}]{shor1995}%
  \BibitemOpen
  \bibfield  {author} {\bibinfo {author} {\bibfnamefont {P.~W.}\ \bibnamefont
  {Shor}},\ }\href {\doibase 10.1103/PhysRevA.52.R2493} {\bibfield  {journal}
  {\bibinfo  {journal} {Phys. Rev. A}\ }\textbf {\bibinfo {volume} {52}},\
  \bibinfo {pages} {R2493} (\bibinfo {year} {1995})}\BibitemShut {NoStop}%
\bibitem [{\citenamefont {Steane}(1996)}]{steane1996}%
  \BibitemOpen
  \bibfield  {author} {\bibinfo {author} {\bibfnamefont {A.}~\bibnamefont
  {Steane}},\ }\href@noop {} {\bibfield  {journal} {\bibinfo  {journal}
  {Physical Review Letters}\ }\textbf {\bibinfo {volume} {77}},\ \bibinfo
  {pages} {793} (\bibinfo {year} {1996})}\BibitemShut {NoStop}%
\bibitem [{\citenamefont {Calderbank}\ and\ \citenamefont
  {Shor}(1996)}]{calderbank1996}%
  \BibitemOpen
  \bibfield  {author} {\bibinfo {author} {\bibfnamefont {A.}~\bibnamefont
  {Calderbank}}\ and\ \bibinfo {author} {\bibfnamefont {P.}~\bibnamefont
  {Shor}},\ }\href@noop {} {\bibfield  {journal} {\bibinfo  {journal} {Physical
  Review A}\ }\textbf {\bibinfo {volume} {54}},\ \bibinfo {pages} {1098}
  (\bibinfo {year} {1996})}\BibitemShut {NoStop}%
\bibitem [{\citenamefont {Aharonov}\ and\ \citenamefont
  {Ben-Or}(1997)}]{aharonov1997}%
  \BibitemOpen
  \bibfield  {author} {\bibinfo {author} {\bibfnamefont {D.}~\bibnamefont
  {Aharonov}}\ and\ \bibinfo {author} {\bibfnamefont {M.}~\bibnamefont
  {Ben-Or}},\ }in\ \href {\doibase 10.1145/258533.258579} {\emph {\bibinfo
  {booktitle} {Proceedings of the Twenty-ninth Annual ACM Symposium on Theory
  of Computing}}},\ \bibinfo {series and number} {STOC '97}\ (\bibinfo
  {publisher} {ACM},\ \bibinfo {address} {New York, NY, USA},\ \bibinfo {year}
  {1997})\ pp.\ \bibinfo {pages} {176--188}\BibitemShut {NoStop}%
\bibitem [{\citenamefont {Knill}\ \emph {et~al.}(1998)\citenamefont {Knill},
  \citenamefont {Laflamme},\ and\ \citenamefont {Zurek}}]{knill1998}%
  \BibitemOpen
  \bibfield  {author} {\bibinfo {author} {\bibfnamefont {E.}~\bibnamefont
  {Knill}}, \bibinfo {author} {\bibfnamefont {R.}~\bibnamefont {Laflamme}}, \
  and\ \bibinfo {author} {\bibfnamefont {W.~H.}\ \bibnamefont {Zurek}},\ }\href
  {\doibase 10.1098/rspa.1998.0166} {\bibfield  {journal} {\bibinfo  {journal}
  {Proceedings of the Royal Society of London A: Mathematical, Physical and
  Engineering Sciences}\ }\textbf {\bibinfo {volume} {454}},\ \bibinfo {pages}
  {365} (\bibinfo {year} {1998})}\BibitemShut {NoStop}%
\bibitem [{\citenamefont {Aliferis}\ \emph {et~al.}(2006)\citenamefont
  {Aliferis}, \citenamefont {Gottesman},\ and\ \citenamefont
  {Preskill}}]{aliferis2006}%
  \BibitemOpen
  \bibfield  {author} {\bibinfo {author} {\bibfnamefont {P.}~\bibnamefont
  {Aliferis}}, \bibinfo {author} {\bibfnamefont {D.}~\bibnamefont {Gottesman}},
  \ and\ \bibinfo {author} {\bibfnamefont {J.}~\bibnamefont {Preskill}},\
  }\href@noop {} {\bibfield  {journal} {\bibinfo  {journal} {Quantum
  Information and Computation}\ }\textbf {\bibinfo {volume} {6}},\ \bibinfo
  {pages} {97} (\bibinfo {year} {2006})}\BibitemShut {NoStop}%
\bibitem [{\citenamefont {Aharonov}\ and\ \citenamefont
  {Ben-Or}(2008)}]{aharonov2008}%
  \BibitemOpen
  \bibfield  {author} {\bibinfo {author} {\bibfnamefont {D.}~\bibnamefont
  {Aharonov}}\ and\ \bibinfo {author} {\bibfnamefont {M.}~\bibnamefont
  {Ben-Or}},\ }\enquote {\bibinfo {title} {Fault-tolerant quantum computation
  with constant error rate},}\ \ (\bibinfo {year} {2008})\ pp.\ \bibinfo
  {pages} {1207--1282}\BibitemShut {NoStop}%
\bibitem [{\citenamefont {Gottesman}(2010)}]{gottesman2009intro}%
  \BibitemOpen
  \bibfield  {author} {\bibinfo {author} {\bibfnamefont {D.}~\bibnamefont
  {Gottesman}},\ }in\ \href {\doibase 10.1090/psapm/068/2762145} {\emph
  {\bibinfo {booktitle} {Quantum Information Science and Its Contributions to
  Mathematics}}}\ (\bibinfo  {publisher} {Amer. Math. Soc.},\ \bibinfo {year}
  {2010})\ pp.\ \bibinfo {pages} {13--58}\BibitemShut {NoStop}%
\bibitem [{\citenamefont {Terhal}(2015)}]{RevModPhys.87.307}%
  \BibitemOpen
  \bibfield  {author} {\bibinfo {author} {\bibfnamefont {B.~M.}\ \bibnamefont
  {Terhal}},\ }\href {\doibase 10.1103/RevModPhys.87.307} {\bibfield  {journal}
  {\bibinfo  {journal} {Rev. Mod. Phys.}\ }\textbf {\bibinfo {volume} {87}},\
  \bibinfo {pages} {307} (\bibinfo {year} {2015})}\BibitemShut {NoStop}%
\bibitem [{\citenamefont {Nielsen}\ and\ \citenamefont
  {Chuang}(2010)}]{nielsen2010}%
  \BibitemOpen
  \bibfield  {author} {\bibinfo {author} {\bibfnamefont {M.~A.}\ \bibnamefont
  {Nielsen}}\ and\ \bibinfo {author} {\bibfnamefont {I.~L.}\ \bibnamefont
  {Chuang}},\ }\href@noop {} {\emph {\bibinfo {title} {Quantum Computation and
  Quantum Information}}},\ \bibinfo {edition} {10th}\ ed.\ (\bibinfo
  {publisher} {Cambridge University Press},\ \bibinfo {year}
  {2010})\BibitemShut {NoStop}%
\bibitem [{\citenamefont {Haake}(2001)}]{haake2001}%
  \BibitemOpen
  \bibfield  {author} {\bibinfo {author} {\bibfnamefont {F.}~\bibnamefont
  {Haake}},\ }\href@noop {} {\emph {\bibinfo {title} {Quantum Signatures of
  Chaos}}},\ \bibinfo {edition} {2nd}\ ed.\ (\bibinfo  {publisher} {Springer},\
  \bibinfo {year} {2001})\BibitemShut {NoStop}%
\bibitem [{\citenamefont {Breuer}\ and\ \citenamefont
  {Petruccione}(2002)}]{breuer2002}%
  \BibitemOpen
  \bibfield  {author} {\bibinfo {author} {\bibfnamefont {H.-P.}\ \bibnamefont
  {Breuer}}\ and\ \bibinfo {author} {\bibfnamefont {F.}~\bibnamefont
  {Petruccione}},\ }\href@noop {} {\emph {\bibinfo {title} {The Theory of Open
  Quantum Systems}}}\ (\bibinfo  {publisher} {Oxford},\ \bibinfo {year}
  {2002})\BibitemShut {NoStop}%
\bibitem [{\citenamefont {Weiss}(2012)}]{weiss2012}%
  \BibitemOpen
  \bibfield  {author} {\bibinfo {author} {\bibfnamefont {U.}~\bibnamefont
  {Weiss}},\ }\href@noop {} {\emph {\bibinfo {title} {Quantum Dissipative
  Systems}}},\ \bibinfo {edition} {4th}\ ed.\ (\bibinfo  {publisher} {World
  Scientific},\ \bibinfo {year} {2012})\BibitemShut {NoStop}%
\bibitem [{\citenamefont {Knill}\ and\ \citenamefont
  {Laflamme}(1996)}]{knill1996}%
  \BibitemOpen
  \bibfield  {author} {\bibinfo {author} {\bibfnamefont {E.}~\bibnamefont
  {Knill}}\ and\ \bibinfo {author} {\bibfnamefont {R.}~\bibnamefont
  {Laflamme}},\ }\href@noop {} {\bibfield  {journal} {\bibinfo  {journal}
  {quant-ph/9608012}\ ,\ \bibinfo {pages} {1}} (\bibinfo {year}
  {1996})}\BibitemShut {NoStop}%
\bibitem [{\citenamefont {Shor}(1996)}]{shor1997}%
  \BibitemOpen
  \bibfield  {author} {\bibinfo {author} {\bibfnamefont {P.~W.}\ \bibnamefont
  {Shor}},\ }in\ \href@noop {} {\emph {\bibinfo {booktitle} {Foundations of
  Computer Science, 1996. Proceedings., 37th Annual Symposium on}}}\ (\bibinfo
  {organization} {IEEE},\ \bibinfo {year} {1996})\ pp.\ \bibinfo {pages}
  {56--65}\BibitemShut {NoStop}%
\bibitem [{\citenamefont {Sanders}\ \emph {et~al.}(2016)\citenamefont
  {Sanders}, \citenamefont {Wallman},\ and\ \citenamefont
  {Sanders}}]{sander2016}%
  \BibitemOpen
  \bibfield  {author} {\bibinfo {author} {\bibfnamefont {Y.~R.}\ \bibnamefont
  {Sanders}}, \bibinfo {author} {\bibfnamefont {J.~J.}\ \bibnamefont
  {Wallman}}, \ and\ \bibinfo {author} {\bibfnamefont {B.~C.}\ \bibnamefont
  {Sanders}},\ }\href {http://stacks.iop.org/1367-2630/18/i=1/a=012002}
  {\bibfield  {journal} {\bibinfo  {journal} {New Journal of Physics}\ }\textbf
  {\bibinfo {volume} {18}},\ \bibinfo {pages} {012002} (\bibinfo {year}
  {2016})}\BibitemShut {NoStop}%
\bibitem [{\citenamefont {Kueng}\ \emph {et~al.}(2016)\citenamefont {Kueng},
  \citenamefont {Long}, \citenamefont {Doherty},\ and\ \citenamefont
  {Flammia}}]{ExpFTThresh}%
  \BibitemOpen
  \bibfield  {author} {\bibinfo {author} {\bibfnamefont {R.}~\bibnamefont
  {Kueng}}, \bibinfo {author} {\bibfnamefont {D.~M.}\ \bibnamefont {Long}},
  \bibinfo {author} {\bibfnamefont {A.~C.}\ \bibnamefont {Doherty}}, \ and\
  \bibinfo {author} {\bibfnamefont {S.~T.}\ \bibnamefont {Flammia}},\ }\href
  {\doibase 10.1103/PhysRevLett.117.170502} {\bibfield  {journal} {\bibinfo
  {journal} {Phys. Rev. Lett.}\ }\textbf {\bibinfo {volume} {117}},\ \bibinfo
  {pages} {170502} (\bibinfo {year} {2016})}\BibitemShut {NoStop}%
\bibitem [{\citenamefont {Guti\'errez}\ \emph {et~al.}(2013)\citenamefont
  {Guti\'errez}, \citenamefont {Svec}, \citenamefont {Vargo},\ and\
  \citenamefont {Brown}}]{gutierrez2013}%
  \BibitemOpen
  \bibfield  {author} {\bibinfo {author} {\bibfnamefont {M.}~\bibnamefont
  {Guti\'errez}}, \bibinfo {author} {\bibfnamefont {L.}~\bibnamefont {Svec}},
  \bibinfo {author} {\bibfnamefont {A.}~\bibnamefont {Vargo}}, \ and\ \bibinfo
  {author} {\bibfnamefont {K.~R.}\ \bibnamefont {Brown}},\ }\href@noop {}
  {\bibfield  {journal} {\bibinfo  {journal} {Physical Review A}\ }\textbf
  {\bibinfo {volume} {87}},\ \bibinfo {pages} {030302(R)} (\bibinfo {year}
  {2013})}\BibitemShut {NoStop}%
\bibitem [{\citenamefont {Geller}\ and\ \citenamefont
  {Zhou}(2013)}]{geller2013}%
  \BibitemOpen
  \bibfield  {author} {\bibinfo {author} {\bibfnamefont {M.~R.}\ \bibnamefont
  {Geller}}\ and\ \bibinfo {author} {\bibfnamefont {Z.}~\bibnamefont {Zhou}},\
  }\href@noop {} {\bibfield  {journal} {\bibinfo  {journal} {Physical Review
  A}\ }\textbf {\bibinfo {volume} {88}},\ \bibinfo {pages} {012314} (\bibinfo
  {year} {2013})}\BibitemShut {NoStop}%
\bibitem [{\citenamefont {Puzzuoli}\ \emph {et~al.}(2014)\citenamefont
  {Puzzuoli}, \citenamefont {Granade}, \citenamefont {Haas}, \citenamefont
  {Criger}, \citenamefont {Magesan},\ and\ \citenamefont
  {Cory}}]{puzzuoli2014}%
  \BibitemOpen
  \bibfield  {author} {\bibinfo {author} {\bibfnamefont {D.}~\bibnamefont
  {Puzzuoli}}, \bibinfo {author} {\bibfnamefont {C.}~\bibnamefont {Granade}},
  \bibinfo {author} {\bibfnamefont {H.}~\bibnamefont {Haas}}, \bibinfo {author}
  {\bibfnamefont {B.}~\bibnamefont {Criger}}, \bibinfo {author} {\bibfnamefont
  {E.}~\bibnamefont {Magesan}}, \ and\ \bibinfo {author} {\bibfnamefont
  {D.~G.}\ \bibnamefont {Cory}},\ }\href@noop {} {\bibfield  {journal}
  {\bibinfo  {journal} {Physical Review A}\ }\textbf {\bibinfo {volume} {89}},\
  \bibinfo {pages} {022306} (\bibinfo {year} {2014})}\BibitemShut {NoStop}%
\bibitem [{\citenamefont {Witzel}\ \emph {et~al.}(2014)\citenamefont {Witzel},
  \citenamefont {Young},\ and\ \citenamefont {Sarma}}]{witzel2014}%
  \BibitemOpen
  \bibfield  {author} {\bibinfo {author} {\bibfnamefont {W.~M.}\ \bibnamefont
  {Witzel}}, \bibinfo {author} {\bibfnamefont {K.}~\bibnamefont {Young}}, \
  and\ \bibinfo {author} {\bibfnamefont {S.~D.}\ \bibnamefont {Sarma}},\
  }\href@noop {} {\bibfield  {journal} {\bibinfo  {journal} {Physical Review
  B}\ }\textbf {\bibinfo {volume} {90}},\ \bibinfo {pages} {115431} (\bibinfo
  {year} {2014})}\BibitemShut {NoStop}%
\bibitem [{\citenamefont {Darmawan}\ and\ \citenamefont
  {Poulin}(2016)}]{darmawan2016tensor}%
  \BibitemOpen
  \bibfield  {author} {\bibinfo {author} {\bibfnamefont {A.~S.}\ \bibnamefont
  {Darmawan}}\ and\ \bibinfo {author} {\bibfnamefont {D.}~\bibnamefont
  {Poulin}},\ }\href@noop {} {\bibfield  {journal} {\bibinfo  {journal} {arXiv
  preprint arXiv:1607.06460}\ } (\bibinfo {year} {2016})}\BibitemShut {NoStop}%
\bibitem [{\citenamefont {Guti\'errez}\ \emph {et~al.}(2016)\citenamefont
  {Guti\'errez}, \citenamefont {Smith}, \citenamefont {Lulushi}, \citenamefont
  {Janardan},\ and\ \citenamefont {Brown}}]{gutierrez2016errors}%
  \BibitemOpen
  \bibfield  {author} {\bibinfo {author} {\bibfnamefont {M.}~\bibnamefont
  {Guti\'errez}}, \bibinfo {author} {\bibfnamefont {C.}~\bibnamefont {Smith}},
  \bibinfo {author} {\bibfnamefont {L.}~\bibnamefont {Lulushi}}, \bibinfo
  {author} {\bibfnamefont {S.}~\bibnamefont {Janardan}}, \ and\ \bibinfo
  {author} {\bibfnamefont {K.~R.}\ \bibnamefont {Brown}},\ }\href {\doibase
  10.1103/PhysRevA.94.042338} {\bibfield  {journal} {\bibinfo  {journal} {Phys.
  Rev. A}\ }\textbf {\bibinfo {volume} {94}},\ \bibinfo {pages} {042338}
  (\bibinfo {year} {2016})}\BibitemShut {NoStop}%
\bibitem [{\citenamefont {Gottesman}(1998)}]{gottesman1998heisenberg}%
  \BibitemOpen
  \bibfield  {author} {\bibinfo {author} {\bibfnamefont {D.}~\bibnamefont
  {Gottesman}},\ }\href@noop {} {\bibfield  {journal} {\bibinfo  {journal}
  {arXiv preprint quant-ph/9807006}\ } (\bibinfo {year} {1998})}\BibitemShut
  {NoStop}%
\bibitem [{\citenamefont {Aaronson}\ and\ \citenamefont
  {Gottesman}(2004)}]{gottesman2004}%
  \BibitemOpen
  \bibfield  {author} {\bibinfo {author} {\bibfnamefont {S.}~\bibnamefont
  {Aaronson}}\ and\ \bibinfo {author} {\bibfnamefont {D.}~\bibnamefont
  {Gottesman}},\ }\href@noop {} {\bibfield  {journal} {\bibinfo  {journal}
  {Physical Review A}\ }\textbf {\bibinfo {volume} {70}},\ \bibinfo {pages}
  {052328} (\bibinfo {year} {2004})}\BibitemShut {NoStop}%
\bibitem [{\citenamefont {Rahn}\ \emph {et~al.}(2002)\citenamefont {Rahn},
  \citenamefont {Doherty},\ and\ \citenamefont {Mabuchi}}]{Rahn2002}%
  \BibitemOpen
  \bibfield  {author} {\bibinfo {author} {\bibfnamefont {B.}~\bibnamefont
  {Rahn}}, \bibinfo {author} {\bibfnamefont {A.~C.}\ \bibnamefont {Doherty}}, \
  and\ \bibinfo {author} {\bibfnamefont {H.}~\bibnamefont {Mabuchi}},\ }\href
  {\doibase 10.1103/PhysRevA.66.032304} {\bibfield  {journal} {\bibinfo
  {journal} {Phys. Rev. A}\ }\textbf {\bibinfo {volume} {66}},\ \bibinfo
  {pages} {032304} (\bibinfo {year} {2002})}\BibitemShut {NoStop}%
\bibitem [{\citenamefont {Bombin}\ and\ \citenamefont
  {Martin-Delgado}(2007)}]{OptimalResourcesTop}%
  \BibitemOpen
  \bibfield  {author} {\bibinfo {author} {\bibfnamefont {H.}~\bibnamefont
  {Bombin}}\ and\ \bibinfo {author} {\bibfnamefont {M.~A.}\ \bibnamefont
  {Martin-Delgado}},\ }\href {\doibase 10.1103/PhysRevA.76.012305} {\bibfield
  {journal} {\bibinfo  {journal} {Phys. Rev. A}\ }\textbf {\bibinfo {volume}
  {76}},\ \bibinfo {pages} {012305} (\bibinfo {year} {2007})}\BibitemShut
  {NoStop}%
\bibitem [{\citenamefont {Horsman}\ \emph {et~al.}(2012)\citenamefont
  {Horsman}, \citenamefont {Fowler}, \citenamefont {Devitt},\ and\
  \citenamefont {Meter}}]{SCLattSurgery}%
  \BibitemOpen
  \bibfield  {author} {\bibinfo {author} {\bibfnamefont {C.}~\bibnamefont
  {Horsman}}, \bibinfo {author} {\bibfnamefont {A.~G.}\ \bibnamefont {Fowler}},
  \bibinfo {author} {\bibfnamefont {S.}~\bibnamefont {Devitt}}, \ and\ \bibinfo
  {author} {\bibfnamefont {R.~V.}\ \bibnamefont {Meter}},\ }\href
  {http://stacks.iop.org/1367-2630/14/i=12/a=123011} {\bibfield  {journal}
  {\bibinfo  {journal} {New Journal of Physics}\ }\textbf {\bibinfo {volume}
  {14}},\ \bibinfo {pages} {123011} (\bibinfo {year} {2012})}\BibitemShut
  {NoStop}%
\bibitem [{\citenamefont {Kitaev}(2003)}]{FTQCAnyons}%
  \BibitemOpen
  \bibfield  {author} {\bibinfo {author} {\bibfnamefont {A.}~\bibnamefont
  {Kitaev}},\ }\href {\doibase http://dx.doi.org/10.1016/S0003-4916(02)00018-0}
  {\bibfield  {journal} {\bibinfo  {journal} {Annals of Physics}\ }\textbf
  {\bibinfo {volume} {303}},\ \bibinfo {pages} {2 } (\bibinfo {year}
  {2003})}\BibitemShut {NoStop}%
\bibitem [{\citenamefont {Cross}\ \emph {et~al.}(2009)\citenamefont {Cross},
  \citenamefont {DiVincenzo},\ and\ \citenamefont {Terhal}}]{cross2009}%
  \BibitemOpen
  \bibfield  {author} {\bibinfo {author} {\bibfnamefont {A.}~\bibnamefont
  {Cross}}, \bibinfo {author} {\bibfnamefont {D.~P.}\ \bibnamefont
  {DiVincenzo}}, \ and\ \bibinfo {author} {\bibfnamefont {B.~M.}\ \bibnamefont
  {Terhal}},\ }\href@noop {} {\bibfield  {journal} {\bibinfo  {journal}
  {quant-ph/0711.1556v2}\ }\textbf {\bibinfo {volume} {x}},\ \bibinfo {pages}
  {1} (\bibinfo {year} {2009})}\BibitemShut {NoStop}%
\bibitem [{\citenamefont {Cohen-Tannoudji}\ \emph {et~al.}(1989)\citenamefont
  {Cohen-Tannoudji}, \citenamefont {Dupont-Roc},\ and\ \citenamefont
  {Grynberg}}]{cohentannoudji1989}%
  \BibitemOpen
  \bibfield  {author} {\bibinfo {author} {\bibfnamefont {C.}~\bibnamefont
  {Cohen-Tannoudji}}, \bibinfo {author} {\bibfnamefont {J.}~\bibnamefont
  {Dupont-Roc}}, \ and\ \bibinfo {author} {\bibfnamefont {G.}~\bibnamefont
  {Grynberg}},\ }\href@noop {} {\emph {\bibinfo {title} {Photons and Atoms}}}\
  (\bibinfo  {publisher} {Wiley},\ \bibinfo {year} {1989})\BibitemShut
  {NoStop}%
\bibitem [{\citenamefont {Heitler}(1954)}]{heitler1954}%
  \BibitemOpen
  \bibfield  {author} {\bibinfo {author} {\bibfnamefont {W.}~\bibnamefont
  {Heitler}},\ }\href@noop {} {\emph {\bibinfo {title} {The Quantum Theory of
  Radiation}}},\ \bibinfo {edition} {3rd}\ ed.\ (\bibinfo  {publisher}
  {Dover},\ \bibinfo {year} {1954})\BibitemShut {NoStop}%
\bibitem [{\citenamefont {Ernst}\ \emph {et~al.}(1987)\citenamefont {Ernst},
  \citenamefont {Bodenhausen},\ and\ \citenamefont {Wokaun}}]{ernst1987}%
  \BibitemOpen
  \bibfield  {author} {\bibinfo {author} {\bibfnamefont {R.}~\bibnamefont
  {Ernst}}, \bibinfo {author} {\bibfnamefont {G.}~\bibnamefont {Bodenhausen}},
  \ and\ \bibinfo {author} {\bibfnamefont {A.}~\bibnamefont {Wokaun}},\
  }\href@noop {} {\emph {\bibinfo {title} {Principles of Nuclear Magnetic
  Resonance in One and Two Dimensions}}}\ (\bibinfo  {publisher} {Oxford},\
  \bibinfo {year} {1987})\BibitemShut {NoStop}%
\bibitem [{\citenamefont {Yariv}(1996)}]{yariv1996}%
  \BibitemOpen
  \bibfield  {author} {\bibinfo {author} {\bibfnamefont {A.}~\bibnamefont
  {Yariv}},\ }\href@noop {} {\emph {\bibinfo {title} {Optical Electronics in
  Modern Communications}}},\ \bibinfo {edition} {5th}\ ed.\ (\bibinfo
  {publisher} {Oxford},\ \bibinfo {year} {1996})\BibitemShut {NoStop}%
\bibitem [{\citenamefont {Mandel}\ and\ \citenamefont
  {Wolf}(1995)}]{mandel1995}%
  \BibitemOpen
  \bibfield  {author} {\bibinfo {author} {\bibfnamefont {L.}~\bibnamefont
  {Mandel}}\ and\ \bibinfo {author} {\bibfnamefont {E.}~\bibnamefont {Wolf}},\
  }\href@noop {} {\emph {\bibinfo {title} {Optical Coherence and Quantum
  Optics}}}\ (\bibinfo  {publisher} {Cambridge University Press},\ \bibinfo
  {year} {1995})\BibitemShut {NoStop}%
\bibitem [{\citenamefont {Barnes}\ and\ \citenamefont
  {Warren}(1999)}]{barnes1999}%
  \BibitemOpen
  \bibfield  {author} {\bibinfo {author} {\bibfnamefont {J.}~\bibnamefont
  {Barnes}}\ and\ \bibinfo {author} {\bibfnamefont {W.}~\bibnamefont
  {Warren}},\ }\href@noop {} {\bibfield  {journal} {\bibinfo  {journal}
  {Physical Review A}\ }\textbf {\bibinfo {volume} {60}},\ \bibinfo {pages}
  {4363} (\bibinfo {year} {1999})}\BibitemShut {NoStop}%
\bibitem [{\citenamefont {Merrill}\ and\ \citenamefont
  {Brown}(2014)}]{merrill2014}%
  \BibitemOpen
  \bibfield  {author} {\bibinfo {author} {\bibfnamefont {J.~T.}\ \bibnamefont
  {Merrill}}\ and\ \bibinfo {author} {\bibfnamefont {K.~R.}\ \bibnamefont
  {Brown}},\ }\enquote {\bibinfo {title} {Progress in compensating pulse
  sequences for quantum computation},}\ in\ \href {\doibase
  10.1002/9781118742631.ch10} {\emph {\bibinfo {booktitle} {Quantum Information
  and Computation for Chemistry}}}\ (\bibinfo  {publisher} {John Wiley and
  Sons},\ \bibinfo {year} {2014})\ pp.\ \bibinfo {pages} {241--294}\BibitemShut
  {NoStop}%
\bibitem [{\citenamefont {Coherent}(2016)}]{coherent2016}%
  \BibitemOpen
  \bibfield  {author} {\bibinfo {author} {\bibnamefont {Coherent}},\ }\href
  {http://www.coherent.com} {\emph {\bibinfo {title} {Coherent Ultrafast Laser
  Systems 2016 Product Catalog}}},\ Vol.~\bibinfo {volume} {0}\ (\bibinfo
  {publisher} {Coherent},\ \bibinfo {year} {2016})\ p.~\bibinfo {pages}
  {0}\BibitemShut {NoStop}%
\bibitem [{\citenamefont {DiVincenzo}\ and\ \citenamefont
  {Aliferis}(2007)}]{divincenzo2007}%
  \BibitemOpen
  \bibfield  {author} {\bibinfo {author} {\bibfnamefont {D.~P.}\ \bibnamefont
  {DiVincenzo}}\ and\ \bibinfo {author} {\bibfnamefont {P.}~\bibnamefont
  {Aliferis}},\ }\href {\doibase 10.1103/PhysRevLett.98.020501} {\bibfield
  {journal} {\bibinfo  {journal} {Phys. Rev. Lett.}\ }\textbf {\bibinfo
  {volume} {98}},\ \bibinfo {pages} {020501} (\bibinfo {year}
  {2007})}\BibitemShut {NoStop}%
\bibitem [{\citenamefont {Svore}\ \emph {et~al.}(2006)\citenamefont {Svore},
  \citenamefont {Cross}, \citenamefont {Chuang},\ and\ \citenamefont
  {Aho}}]{svore2006}%
  \BibitemOpen
  \bibfield  {author} {\bibinfo {author} {\bibfnamefont {K.~M.}\ \bibnamefont
  {Svore}}, \bibinfo {author} {\bibfnamefont {A.~W.}\ \bibnamefont {Cross}},
  \bibinfo {author} {\bibfnamefont {I.~L.}\ \bibnamefont {Chuang}}, \ and\
  \bibinfo {author} {\bibfnamefont {A.~V.}\ \bibnamefont {Aho}},\ }\href@noop
  {} {\bibfield  {journal} {\bibinfo  {journal} {Quantum Information and
  Computation}\ }\textbf {\bibinfo {volume} {6}},\ \bibinfo {pages} {193}
  (\bibinfo {year} {2006})}\BibitemShut {NoStop}%
\bibitem [{\citenamefont {Tomita}\ and\ \citenamefont
  {Svore}(2014)}]{LowDSCRealNoise}%
  \BibitemOpen
  \bibfield  {author} {\bibinfo {author} {\bibfnamefont {Y.}~\bibnamefont
  {Tomita}}\ and\ \bibinfo {author} {\bibfnamefont {K.~M.}\ \bibnamefont
  {Svore}},\ }\href {\doibase 10.1103/PhysRevA.90.062320} {\bibfield  {journal}
  {\bibinfo  {journal} {Phys. Rev. A}\ }\textbf {\bibinfo {volume} {90}},\
  \bibinfo {pages} {062320} (\bibinfo {year} {2014})}\BibitemShut {NoStop}%
\bibitem [{\citenamefont {Fowler}\ \emph {et~al.}(2012)\citenamefont {Fowler},
  \citenamefont {Whiteside},\ and\ \citenamefont {Hollenberg}}]{MWPMSC}%
  \BibitemOpen
  \bibfield  {author} {\bibinfo {author} {\bibfnamefont {A.~G.}\ \bibnamefont
  {Fowler}}, \bibinfo {author} {\bibfnamefont {A.~C.}\ \bibnamefont
  {Whiteside}}, \ and\ \bibinfo {author} {\bibfnamefont {L.~C.~L.}\
  \bibnamefont {Hollenberg}},\ }\href {\doibase 10.1103/PhysRevLett.108.180501}
  {\bibfield  {journal} {\bibinfo  {journal} {Phys. Rev. Lett.}\ }\textbf
  {\bibinfo {volume} {108}},\ \bibinfo {pages} {180501} (\bibinfo {year}
  {2012})}\BibitemShut {NoStop}%
\bibitem [{\citenamefont {Edmonds}(1965{\natexlab{a}})}]{EdmondsOrig}%
  \BibitemOpen
  \bibfield  {author} {\bibinfo {author} {\bibfnamefont {J.}~\bibnamefont
  {Edmonds}},\ }\href@noop {} {\bibfield  {journal} {\bibinfo  {journal}
  {Canad. J. Math.}\ }\textbf {\bibinfo {volume} {17}},\ \bibinfo {pages} {449}
  (\bibinfo {year} {1965}{\natexlab{a}})}\BibitemShut {NoStop}%
\bibitem [{\citenamefont {Edmonds}(1965{\natexlab{b}})}]{EdmondsAlgo}%
  \BibitemOpen
  \bibfield  {author} {\bibinfo {author} {\bibfnamefont {J.}~\bibnamefont
  {Edmonds}},\ }\href@noop {} {\bibfield  {journal} {\bibinfo  {journal} {J.
  Res. Nat. Bur. Standards}\ }\textbf {\bibinfo {volume} {69B}},\ \bibinfo
  {pages} {125} (\bibinfo {year} {1965}{\natexlab{b}})}\BibitemShut {NoStop}%
\bibitem [{\citenamefont {Wallman}\ and\ \citenamefont
  {Emerson}(2016)}]{PhysRevA.94.052325}%
  \BibitemOpen
  \bibfield  {author} {\bibinfo {author} {\bibfnamefont {J.~J.}\ \bibnamefont
  {Wallman}}\ and\ \bibinfo {author} {\bibfnamefont {J.}~\bibnamefont
  {Emerson}},\ }\href {\doibase 10.1103/PhysRevA.94.052325} {\bibfield
  {journal} {\bibinfo  {journal} {Phys. Rev. A}\ }\textbf {\bibinfo {volume}
  {94}},\ \bibinfo {pages} {052325} (\bibinfo {year} {2016})}\BibitemShut
  {NoStop}%
\end{thebibliography}
\end{document}